\documentclass[preprint,authoryear,11pt]{elsarticle}

\usepackage{graphicx,subcaption}
\usepackage{epstopdf}
\usepackage{amsmath, amssymb}
\usepackage{rotating}
\usepackage{morefloats}
\usepackage{comment}
\usepackage{natbib}
\usepackage{caption}
\usepackage{booktabs}
\usepackage{multirow}
\usepackage{longtable}
\usepackage{setspace}
\usepackage{tikz}
\usepackage{pgfplots}
\pgfplotsset{compat=1.18}
\usepackage[margin=2.5cm]{geometry}
\usepackage{bbm}
\usepackage{algorithm}
\usepackage{algpseudocode}
\usepackage{pdflscape}
\usepackage{amsthm}

\theoremstyle{remark}
\newtheorem{rmk}{Remark}

\newcommand{\w}{\boldsymbol{w}}
\newcommand{\wt}{\w^{\!\top}}
\newcommand{\tw}{\boldsymbol{\tilde{w}}}
\newcommand{\twt}{\boldsymbol{\tilde{w}}^{\!\top}}

\begin{document}
\title{Entropic Value-at-Risk portfolio optimization for tempered stable L\'evy processes}
\author{Jaehyung Choi\corref{cor}}
\ead{jj.jaehyung.choi@gmail.com}
\cortext[cor]{Corresponding author}

\begin{abstract}
	We develop parametric Entropic Value-at-Risk (EVaR) portfolio optimization for tempered stable L\'evy returns. We derive portfolio cumulant-generating functions and weight-dependent admissible moment-generating-function domains under two multivariate constructions: a multivariate normal tempered stable approach and an independent component factorization. These expressions allow portfolio EVaR to be evaluated from fitted asset- or component-level parameters without repeated portfolio-level distribution fitting. We construct minimum-EVaR portfolios and two entropic reward--risk portfolios. We test the portfolios in a rolling 2000 to 2026 out-of-sample U.S. sector ETF allocation. In this universe, several entropic portfolios have higher realized Sharpe ratios than their matched CVaR portfolios or standard allocation benchmarks.
\end{abstract}
\maketitle

\section{Introduction}
	Portfolio construction is a central topic in both academic finance and investment management practice. \cite{markowitz1952portfolio} established the mean--variance framework, casting portfolio selection as an optimization problem that balances expected return against variance. Despite its profound impact, the mean--variance framework has several notable shortcomings. A well-known limitation of the mean--variance optimization is its sensitivity to estimated inputs, such as expected returns and the covariance matrix (\cite{michaud1989markowitz}). Due to this sensitivity, the resulting portfolio can change substantially as estimated inputs vary with volatile market conditions, particularly when return distributions exhibit skewness and heavy tails. Existing remedies such as shrinkage estimation (\cite{ledoit2003improved}), regularization (\cite{demiguel2009generalized}), and portfolio constraints (\cite{jagannathan2003risk}) mitigate estimation sensitivity but do not eliminate it. More fundamentally, the Markowitz framework adopts variance as a proxy for risk, which captures total return variability rather than downside risk.

	Among alternative downside risk measures, Value-at-Risk (VaR) and Conditional Value-at-Risk (CVaR) (\cite{rockafellar2000optimization, rockafellar2002conditional}) have become widely used in portfolio optimization. VaR focuses explicitly on downside quantiles rather than total return variability. However, it does not capture the severity of losses beyond the VaR threshold and is not coherent in general (\cite{artzner1999coherent}). CVaR, also referred to as Expected Shortfall, addresses these limitations by measuring the average loss beyond the VaR threshold and retains desirable properties for portfolio optimization. Both measures consider only particular parts of the loss distribution.
	
	A more recent alternative is Entropic Value-at-Risk (EVaR), an information-theoretic extension of VaR based on the Chernoff inequality (\cite{ahmadi2012entropic}). EVaR is a coherent risk measure and provides an upper bound for both VaR and CVaR. Its definition uses the moment-generating function of the return distribution, and EVaR also has a dual representation based on information-theoretic divergences such as the Kullback--Leibler divergence. However, EVaR-based portfolio optimization encounters two technical issues. First, the moment-generating function required by EVaR does not exist for all return distributions. Second, even when it exists, the parameters and admissible domain of the portfolio return distribution generally vary with the portfolio weights unless the projected distribution can be determined directly from fitted asset- or factor-level parameters.
	
	Existing EVaR studies under L\'evy models have focused mainly on EVaR evaluation for a specified return or loss process. \cite{mishura2024properties} studies EVaR for selected L\'evy classes, and \cite{nedeltchev2026measuring} characterizes the scalar EVaR problem for several parametric market-risk models. These studies consider scalar EVaR but not the multivariate setting in which the projected cumulant and its admissible domain vary with the portfolio weights. Tempered stable processes are relevant here because they have been used to model skewness and heavy tails in quantitative finance (\cite{cont2003financial, rachev2011financial, kim2023deep, kim2023multi, tsuchida2012mean, beck2013empirical, choi2015reward, choi2025diversified, georgiev2015periodic, anand2017equity}).
	
	\cite{ahmadi2019portfolio} and \cite{cajas2021entropic} develop convex formulations for sample-based or finite-scenario EVaR portfolio optimization. \cite{luxenberg2024portfolio} obtains a convex minimum-EVaR formulation for Gaussian mixture returns using a reciprocal transformation and a cumulant perspective. \cite{firouzi2014optimal} considers parametric EVaR optimization under jump-diffusion models. \cite{hitaj2015portfolio} combines Independent Component Analysis (ICA) with Mixed Tempered Stable distributions for modified VaR and Expected Shortfall. However, these studies do not consider the multivariate tempered stable EVaR problem studied here.

	In this paper, we develop EVaR portfolio optimization for tempered stable L\'evy returns. First, we derive the portfolio cumulant and the weight-dependent admissible moment-generating-function domain under MNTS projection and independent component factorization with tempered stable marginals. These expressions allow portfolio EVaR to be evaluated across candidate weights without refitting the portfolio return distribution. Second, we construct minimum-EVaR portfolios and two entropic reward--risk portfolios. Third, we compare the portfolios out of sample in a rolling U.S. sector ETF allocation across risk measures, multivariate representations, tail families, and portfolio objectives.
	
	The remainder of this paper is structured as follows. Section~\ref{sec_prelim} reviews the risk measures, reward--risk ratios, and L\'evy models used in the analysis. Section~\ref{sec_levy_evar} develops the cumulant representation of EVaR and its numerical evaluation. Section~\ref{sec_ica_evar} develops two portfolio EVaR approaches and the entropic portfolio problems. Section~\ref{sec_result} presents the rolling backtest results. Section~\ref{sec_conclusion} concludes the paper.

\section{Preliminaries}
\label{sec_prelim}
	This section reviews the risk measures, reward--risk ratios, and L\'evy models used in this paper.
	
\subsection{Risk measures}
	A risk measure \(\mathcal{D}(X)\) assigns a real number to a portfolio series \(X\) to summarize its risk. In this paper, \(X\) and \(L\) denote the return and loss series, respectively, with \(L=-X\).

\subsubsection{VaR and CVaR} 
	VaR is widely used because of its intuitive interpretation as a loss threshold. The VaR at the confidence level of \((1-\eta)100\%\) \((0<\eta<1)\) is defined as the smallest loss threshold \(l\) such that the probability of a portfolio loss exceeding \(l\) does not exceed \(\eta\):
	\begin{align}
	\label{eq_var}
	\mathrm{VaR}_{1-\eta}(X)=\inf\left\{l\in\mathbb{R}\ \Big|\ \Pr(X<-l)\le\eta\right\}.
	\end{align}

	VaR does not measure the magnitude of losses beyond the threshold and is not coherent in general because subadditivity can fail (\cite{artzner1999coherent}).

	CVaR (\cite{rockafellar2000optimization}) accounts for losses beyond the VaR threshold. The CVaR at the confidence level of \((1-\eta)100\%\) is defined as
	\begin{align}
	\label{eq_cvar}
		\textrm{CVaR}_{1-\eta}(X) = \frac{1}{\eta} \int_0^\eta \textrm{VaR}_{1-\zeta}(X) \ d\zeta,
	\end{align}
	where \(0<\eta<1\).
	
	CVaR satisfies the axioms for coherent risk measures, including subadditivity, monotonicity, translation invariance, and positive homogeneity (\cite{rockafellar2000optimization}). By averaging losses beyond the VaR threshold, CVaR captures tail severity that VaR misses.
	
\subsubsection{EVaR}
	\cite{ahmadi2012entropic} defines EVaR using the moment-generating function of a loss distribution \(L\). The EVaR at the confidence level of \((1-\eta)100\%\) (\(0<\eta<1\)) for the loss \(L\) is defined as
	\begin{align}
	\label{eq_evar_loss}
		\textrm{EVaR}_{1-\eta}(L) = \inf_{u\in \mathcal{U}} \left\{ \frac{1}{u} \ln \bigg(\frac{M_{L}(u)}{\eta}\bigg) \right\}, 
	\end{align}
	where \(M_{L}(u)\) is the moment-generating function of the loss distribution. In \cite{ahmadi2012entropic}, \(\mathcal{U}\) is defined as \((0,\infty)\). For the distributions considered in this paper, \(\mathcal{U}\) is the intersection of \((0,\infty)\) and the domain on which the moment-generating function is finite.
	
	In this paper, the parameters of all distributions are estimated on the return time series \(X\) rather than on the loss time series \(L\). Based on the relationship between return and loss given above, the moment-generating function of the loss follows from that of the return after replacing \(u\) with \(-u\), that is, \(M_{L}(u) = M_{X}(-u)\). The EVaR at the confidence level of \((1-\eta)100\%\) (\(0<\eta<1\)) for return time series \(X\) is defined as
	\begin{align}
	\label{eq_evar}
		\textrm{EVaR}_{1-\eta}(X) = \inf_{u\in \mathcal{U}} \left\{ \frac{1}{u} \ln \bigg(\frac{M_{X}(-u)}{\eta}\bigg) \right\}.
	\end{align}
	All EVaR formulas in the following sections use return-based parameters under this sign change. The same estimated parameters can then be used without refitting the sign-inverted returns.
	
	EVaR has several useful properties (\cite{ahmadi2012entropic}). First, EVaR is coherent. Second, it provides upper bounds for VaR and CVaR. Third, the moment-generating function can reflect asymmetry and heavy tails when these features are included in the return distribution. In portfolio optimization, however, both the admissible set \(\mathcal{U}\) and the cumulant of the portfolio return may depend on the portfolio weights.

\subsection{Reward--risk ratios}
	A reward--risk ratio compares a reward functional with a risk functional. We also use reward--risk ratios as objective functions for portfolio optimization. A general reward--risk ratio is given by
	\begin{align}
		RR(X)=\frac{\mathcal{G}(X)}{\mathcal{D}(X)},
	\end{align}
	where \(\mathcal{G}(X)\) denotes a reward functional and \(\mathcal{D}(X)\) denotes a risk functional. 
	
	Table~\ref{tab_reward_risk} summarizes the classical reward--risk ratios considered in this paper. As a numerator, the Sharpe and Stable Tail Adjusted Return (STAR) ratios use expected return, and the Calmar ratio may choose cumulative return, the compound annual growth rate (CAGR), or expected return. The Rachev ratio adopts the lower-tail risk of the sign-inverted returns, i.e., an upper-tail reward of the original return series. For a denominator, the Sharpe ratio penalizes total return volatility regardless of direction, which may understate the attractiveness of right-skewed return distributions. The Calmar ratio compares the selected return measure with the worst peak-to-trough decline, i.e., maximum drawdown (MDD), capturing path-dependent downside risk. The STAR ratio uses CVaR instead of volatility to measure tail losses directly. The Rachev ratio compares upper-tail reward with lower-tail risk by using CVaR in both the numerator and denominator. 

\begin{table}[t]
\centering
\caption{Classical reward--risk ratios as ratios of reward and risk functionals.}
\label{tab_reward_risk}
\begin{tabular}{llll}
\hline
Objective & Reward functional \(\mathcal{G}(X)\) & Risk functional \(\mathcal{D}(X)\) & Reference \\
\hline
Sharpe & \(\mathbb{E}[X]\) & \(\sigma(X)\) & \cite{sharpe1964capital} \\
Calmar & Cumulative return, \(\mathrm{CAGR}(X)\), or \(\mathbb{E}[X]\) & \(\mathrm{MDD}(X)\) & \cite{young1991calmar} \\
STAR & \(\mathbb{E}[X]\) & \(\mathrm{CVaR}_{1-\eta}(X)\) & \cite{martin2003phi} \\
Rachev & \(\mathrm{CVaR}_{1-\zeta}(-X)\) & \(\mathrm{CVaR}_{1-\eta}(X)\) & \cite{biglova2004different} \\
\hline
\end{tabular}
\end{table}
	
\subsection{L\'evy processes}
	A L\'evy process extends Brownian motion by allowing jumps (\cite{sato1999levy}). L\'evy processes and their subclasses are used in financial modeling to describe skewness and heavy tails in asset returns (\cite{rachev2011financial}). 
	
	We introduce L\'evy processes in the multivariate setting and obtain the univariate case by setting \(N=1\). Throughout the theoretical sections, bold italic symbols such as \(\boldsymbol{X}_t\) and \(\boldsymbol{Z}\) denote random vectors, plain italic symbols denote scalar random variables, and bold upright symbols denote observed data matrices or vectors. We denote the in-sample return matrix by \(\mathbf{X}=(\mathbf{x}_1,\ldots,\mathbf{x}_T)\in\mathbb{R}^{N\times T}\), where each column \(\mathbf{x}_t=(x_t^{(1)},\ldots,x_t^{(N)})^{\top}\in\mathbb{R}^{N}\) is an observed realization of the corresponding random return vector \(\boldsymbol{X}_t\).
	
	An \(N\)-dimensional stochastic process \((\boldsymbol{X}_t, \mathbb{P})_{t\in[0,\tau]}\), with \(\tau>0\), is a L\'evy process if it satisfies the following properties: (i) \(\boldsymbol{X}_0 = \boldsymbol{0}\) almost surely; (ii) it has independent increments, such that for any \(0 \le t_0 < t_1 < \dots < t_n\), the increments \(\boldsymbol{X}_{t_1} - \boldsymbol{X}_{t_0}, \dots, \boldsymbol{X}_{t_n} - \boldsymbol{X}_{t_{n-1}}\) are mutually independent; (iii) it has stationary increments, such that \(\boldsymbol{X}_{t+s} - \boldsymbol{X}_s\) has the same distribution as \(\boldsymbol{X}_t\) for all \(s,t\ge0\); and (iv) it is stochastically continuous, i.e., \(\lim_{h \to 0} \Pr(\|\boldsymbol{X}_{t+h}-\boldsymbol{X}_t\|>\epsilon)=0\) for all \(\epsilon>0\). A L\'evy process is described by the triplet \((\boldsymbol{\Sigma}, \nu, \boldsymbol{\mu})\), where \(\boldsymbol{\mu}\in\mathbb{R}^N\) is the drift vector, \(\boldsymbol{\Sigma}\in\mathbb{R}^{N\times N}\) is a symmetric positive semi-definite covariance matrix of the diffusion component, and \(\nu\) is the L\'evy measure governing the jump component.
	
	From the L\'evy--Khintchine formula, the characteristic function of \(\boldsymbol{X}_t\) is given by
	\begin{align}
	\label{charac_ftn_levy}
		\phi_{\boldsymbol{X}_t}(\boldsymbol{z}) = \mathbb{E}[{\rm e}^{i\boldsymbol{z}^{\top}\boldsymbol{X}_t}] = \exp\left(t\Big[-\frac{1}{2}\boldsymbol{z}^{\top}\boldsymbol{\Sigma}\boldsymbol{z} + i \boldsymbol{\mu}^{\top}\boldsymbol{z} + \int_{\mathbb{R}^N} ({\rm e}^{i\boldsymbol{z}^{\top}\boldsymbol{x}} - 1 - i\boldsymbol{z}^{\top}\boldsymbol{x} \mathbbm{1}_{\|\boldsymbol{x}\| \leq 1}) \nu(d\boldsymbol{x})\Big]\right),
	\end{align}
	where \(\|\boldsymbol{x}\| = \sqrt{\boldsymbol{x}^{\top}\boldsymbol{x}}\) and \(\mathbbm{1}_A\) is the indicator function such that \(\mathbbm{1}_A(\boldsymbol{x})=1\) when \(\boldsymbol{x}\in A\) and 0 otherwise. The logarithmic characteristic function \(\Phi_{\boldsymbol{X}_t}(\boldsymbol{z})\), or L\'evy exponent, is given by
	\begin{align}
	\label{log_charac_ftn_levy}
		\Phi_{\boldsymbol{X}_t}(\boldsymbol{z}) = \log \phi_{\boldsymbol{X}_t}(\boldsymbol{z}) = t\Big[-\frac{1}{2}\boldsymbol{z}^{\top}\boldsymbol{\Sigma}\boldsymbol{z} + i \boldsymbol{\mu}^{\top}\boldsymbol{z} + \int_{\mathbb{R}^N} ({\rm e}^{i\boldsymbol{z}^{\top}\boldsymbol{x}} - 1 - i\boldsymbol{z}^{\top}\boldsymbol{x} \mathbbm{1}_{\|\boldsymbol{x}\| \leq 1}) \nu(d\boldsymbol{x})\Big].
	\end{align}

	When finite, the moment-generating function of \(\boldsymbol{X}_t\) is related to its characteristic function \(\phi_{\boldsymbol{X}_t}(\boldsymbol{z})\) by
	\begin{align}
	\label{mgf}
		M_{\boldsymbol{X}_t}(\boldsymbol{u})= \mathbb{E}[e^{\boldsymbol{u}^{\top}\boldsymbol{X}_t}]=\phi_{\boldsymbol{X}_t}(\boldsymbol{z})|_{\boldsymbol{z}=-i\boldsymbol{u}}.
	\end{align}
	When \(N=1\), the L\'evy triplet becomes \((\sigma^2, \nu, \mu)\), where \(\sigma\geq 0\) is the diffusion coefficient, \(\nu\) is the L\'evy measure, and \(\mu\) is the scalar drift. Eqs.~(\ref{charac_ftn_levy})--(\ref{mgf}) reduce to their univariate counterparts.
		
\subsubsection{Normal distribution}
	The normal distribution can be obtained from a L\'evy process with L\'evy triplet \((\boldsymbol{\Sigma}, 0, \boldsymbol{\mu})\), i.e., the Normal case has no jump component. From Eq.~(\ref{charac_ftn_levy}) and Eq.~(\ref{log_charac_ftn_levy}), the characteristic function and the corresponding L\'evy exponent of \((\boldsymbol{X}_t, \mathbb{P})_{t\in[0,\tau]}\) are given by
	\begin{align}
	\label{charac_ftn_normal}
		\phi_{\boldsymbol{X}^{\text{Normal}}_t}(\boldsymbol{z}) &= \exp\left(t\Big[-\frac{1}{2}\boldsymbol{z}^{\top}\boldsymbol{\Sigma}\boldsymbol{z} + i \boldsymbol{\mu}^{\top}\boldsymbol{z} \Big] \right),\\
	\label{log_charac_ftn_normal}
		\Phi_{\boldsymbol{X}^{\text{Normal}}_t}(\boldsymbol{z}) &= t\Big[-\frac{1}{2}\boldsymbol{z}^{\top}\boldsymbol{\Sigma}\boldsymbol{z} + i \boldsymbol{\mu}^{\top}\boldsymbol{z}\Big].
	\end{align}
	When \(N=1\), Eq.~(\ref{charac_ftn_normal}) and Eq.~(\ref{log_charac_ftn_normal}) reduce to those of the univariate normal case with \(\boldsymbol{\Sigma}=\sigma^2\) and \(\boldsymbol{\mu}=\mu\). When \(t=1\), the formulas specialize to those of the normal distribution.
	
\subsubsection{Classical tempered stable process/distribution}
\label{sec_cts_dist}
	A classical tempered stable (CTS) process is a special case of a L\'evy process with L\'evy triplet \((0, \nu, \mu)\). CTS processes can describe heavy-tailed asset returns and have been used in portfolio management, risk management, and option pricing (\cite{tsuchida2012mean, beck2013empirical, choi2015reward, choi2025diversified, georgiev2015periodic, anand2017equity, kim2023deep}).
	
	A CTS distribution \(\mathrm{CTS}(\alpha, C, \lambda_+, \lambda_-, \mu)\) is parameterized by the tail index \(\alpha\), the scale parameter \(C\), the decay parameters \(\lambda_\pm\), and the location parameter \(\mu\) (\cite{rosinski2007tempering}). According to \cite{rachev2011financial}, its L\'evy measure is represented by
	\begin{align}
	\label{levy_measure_cts}
		\nu^{\text{CTS}}(dx;\boldsymbol{\xi})=C\Big(\frac{{\rm e}^{-\lambda_{+}x}}{x^{\alpha+1}} \mathbbm{1}_{x>0}(x)+\frac{{\rm e}^{-\lambda_{-} |x|}}{|x|^{\alpha+1}} \mathbbm{1}_{x<0}(x)\Big)dx,
	\end{align}
	where \(\alpha \in (0,2) \setminus \{1\}\) and \(C, \lambda_+, \lambda_->0\).
	
	Applying the L\'evy measure of Eq.~(\ref{levy_measure_cts}) to Eq.~(\ref{charac_ftn_levy}) and Eq.~(\ref{log_charac_ftn_levy}) with \(N=1\), the characteristic function and the L\'evy exponent of a CTS process are given by
	\begin{align}
	\label{charac_ftn_cts}
	\phi_{X^{\text{CTS}}_t}(z) &= \exp \Bigg\{ i t \Big[\mu - C \Gamma(1-\alpha) (\lambda_+^{\alpha-1} - \lambda_-^{\alpha-1}) \Big] z+ tC \Gamma(-\alpha) \Big[ (\lambda_+ - i z)^\alpha - \lambda_+^\alpha + (\lambda_- + i z)^\alpha - \lambda_-^\alpha \Big] \Bigg\},\\
	\label{log_charac_ftn_cts}
	\Phi_{X^{\text{CTS}}_t}(z) &= i t \Big[\mu -  C \Gamma(1-\alpha) (\lambda_+^{\alpha-1} - \lambda_-^{\alpha-1}) \Big] z + tC \Gamma(-\alpha) \Big[ (\lambda_+ - i z)^\alpha - \lambda_+^\alpha + (\lambda_- + i z)^\alpha - \lambda_-^\alpha \Big].
	\end{align}
	Similarly, when \(t=1\), the expressions reduce to the characteristic function and the L\'evy exponent for a CTS distribution.
	
\subsubsection{Normal tempered stable process/distribution}
	As with CTS processes, normal tempered stable (NTS) processes are pure-jump L\'evy processes with triplet \((0, \nu, \mu)\), where \(\nu\) is the L\'evy measure induced by the NTS distribution. The NTS distribution (\cite{barndorff2002normal}) is obtained by time-changing a normal distribution using a tempered stable subordinator. NTS processes allow asymmetric heavy tails and have been used in portfolio and risk management and in financial time-series models, including ARMA(1,1)--GARCH(1,1)--NTS specifications (\cite{anand2016foster, anand2017equity, kurosaki2019foster}).
	
	For a univariate NTS random variable, i.e., \(X \sim \mathrm{NTS}(\alpha, \theta, \beta, \gamma, \mu)\), \(\alpha\), \(\beta\), \(\theta\), \(\gamma\), and \(\mu\) control tail thickness, skewness, scale, diffusion, and location, respectively. The L\'evy measure of NTS processes (\cite{rachev2011financial}) is represented by
	\begin{align}
	\label{levy_measure_nts}
		\nu^{\text{NTS}}(dx;\boldsymbol{\xi})=\frac{\theta^{1-\frac{\alpha}{2}}}{\Gamma(1-\frac{\alpha}{2})\sqrt{\pi}\gamma}\exp\left(\frac{\beta}{\gamma^{2}}x\right)\left(\frac{|x|}{\sqrt{\beta^{2}+2\theta\gamma^{2}}}\right)^{-\frac{\alpha+1}{2}}
K_{\frac{\alpha+1}{2}}\left(\frac{\sqrt{\beta^{2}+2\theta\gamma^{2}}}{\gamma^{2}}|x|\right)dx,
	\end{align}
	where \(\theta\) and \(\gamma\) are positive, \(\alpha \in (0,2)\), and \(K\) is the modified Bessel function of the second kind.
	
	The characteristic function and the L\'evy exponent of NTS processes follow from inserting Eq.~(\ref{levy_measure_nts}) into Eq.~(\ref{charac_ftn_levy}) and Eq.~(\ref{log_charac_ftn_levy}) with \(N=1\):
	\begin{align}
	\label{charac_ftn_nts}
		\phi_{X^{\text{NTS}}_t}(z)&=\exp \Bigg\{ i t (\mu - \beta) z - \frac{2t \theta}{\alpha} \Big[ (1 - i \frac{\beta}{\theta} z + \frac{\gamma^2}{2\theta} z^2)^{\alpha/2} - 1 \Big] \Bigg\},\\
		\label{log_charac_ftn_nts}
		\Phi_{X^{\text{NTS}}_t}(z)&=i t (\mu - \beta) z - \frac{2t \theta}{\alpha} \Big[ (1 - i \frac{\beta}{\theta} z + \frac{\gamma^2}{2\theta} z^2)^{\alpha/2} - 1 \Big].
	\end{align}
	When \(t=1\), the characteristic function and L\'evy exponent coincide with those of the NTS distribution.

\subsubsection{Multivariate normal tempered stable process/distribution}
\label{sec_mnts_dist}
	The multivariate normal tempered stable (MNTS) distribution (\cite{kim2012measuring, kim2022portfolio, kim2023multi}) extends the univariate NTS construction by time-changing a multivariate Brownian motion with a tempered stable subordinator. Let \((\mathcal{T}_t)_{t\ge0}\) denote the tempered stable subordinator parameterized by \((\alpha,\theta)\) and normalized so that \(\mathbb{E}[\mathcal{T}_t]=t\). Let \((\boldsymbol{B}_t)_{t\ge0}\) be an \(N\)-dimensional Brownian motion, independent of \((\mathcal{T}_t)_{t\ge0}\), with \(\mathrm{Cov}(\boldsymbol{B}_t)=t\boldsymbol{\rho}\) for a correlation matrix \(\boldsymbol{\rho}\). The MNTS process \(\boldsymbol{X}^{\text{MNTS}}_t \sim \mathrm{MNTS}(\alpha, \theta, \boldsymbol{\beta}, \boldsymbol{\gamma}, \boldsymbol{\mu}, \boldsymbol{\rho})\) is defined as
	\begin{align}
	\label{mnts_def}
	\boldsymbol{X}^{\text{MNTS}}_t = \boldsymbol{\mu} t + \boldsymbol{\beta}(\mathcal{T}_t-t)+\mathrm{diag}(\boldsymbol{\gamma})\boldsymbol{B}_{\mathcal{T}_t}.
	\end{align}
	For each fixed \(t\), \(\boldsymbol{B}_{\mathcal{T}_t}\) conditional on \(\mathcal{T}_t\) is Gaussian with covariance \(\mathcal{T}_t\boldsymbol{\rho}\). Equivalently, \(\boldsymbol{B}_{\mathcal{T}_t}\overset{d}{=}\sqrt{\mathcal{T}_t}\boldsymbol{Z}\) for \(\boldsymbol{Z}\sim\mathcal{N}(\boldsymbol{0},\boldsymbol{\rho})\) independent of \(\mathcal{T}_t\). This is a fixed-time normal-mixture representation and does not use a single Gaussian vector to define the entire process. The vectors \(\boldsymbol{\mu}\in\mathbb{R}^N\), \(\boldsymbol{\beta}\in\mathbb{R}^N\), and \(\boldsymbol{\gamma}\in\mathbb{R}_{+}^N\) control the location, asymmetry, and diffusion scale of each marginal, respectively. The parameters \((\alpha,\theta)\) are shared across all marginals through the common subordinator. Cross-marginal dependence is induced by both the common time change and \(\boldsymbol{\rho}\). Each marginal \(X^{\text{MNTS},(i)}_t\) follows the univariate NTS process \(\mathrm{NTS}(\alpha,\theta,\beta_i,\gamma_i,\mu_i)\).

	The characteristic function of the MNTS process is given by
	\begin{align}
	\label{charac_ftn_mnts}
	\phi_{\boldsymbol{X}^{\text{MNTS}}_t}(\boldsymbol{z}) &= \exp \Bigg\{ i t (\boldsymbol{\mu} - \boldsymbol{\beta})^{\top} \boldsymbol{z} - \frac{2 t\theta}{\alpha} \Big[ (1 - i \frac{\boldsymbol{\beta}^{\top} \boldsymbol{z}}{\theta}  + \frac{\boldsymbol{z}^{\top} \boldsymbol{\Sigma} \boldsymbol{z}}{2\theta} )^{\alpha/2} - 1 \Big] \Bigg\},\\
	\label{log_charac_ftn_mnts}
	\Phi_{\boldsymbol{X}^{\text{MNTS}}_t}(\boldsymbol{z}) &= i t (\boldsymbol{\mu} - \boldsymbol{\beta})^{\top} \boldsymbol{z} - \frac{2 t \theta}{\alpha} \Big[ (1 - i \frac{\boldsymbol{\beta}^{\top} \boldsymbol{z}}{\theta}  + \frac{\boldsymbol{z}^{\top} \boldsymbol{\Sigma} \boldsymbol{z}}{2\theta})^{\alpha/2} - 1 \Big],
	\end{align}
	where \(\boldsymbol{\Sigma}=\mathrm{diag}(\boldsymbol{\gamma})\boldsymbol{\rho}\,\mathrm{diag}(\boldsymbol{\gamma})\). When \(N=1\), Eq.~(\ref{charac_ftn_mnts}) and Eq.~(\ref{log_charac_ftn_mnts}) specialize to Eq.~(\ref{charac_ftn_nts}) and Eq.~(\ref{log_charac_ftn_nts}) of the univariate NTS process, respectively. 
	
\section{Entropic Value-at-Risk for L\'evy processes}
\label{sec_levy_evar}
	This section derives EVaR for L\'evy and tempered stable processes. As noted in Section~\ref{sec_prelim}, EVaR requires a finite moment-generating function. We consider L\'evy processes whose moment-generating functions are finite on non-empty parameter-dependent domains. The multivariate case is presented first, and the univariate case follows by setting \(N=1\).

\subsection{General formula}
\label{sec_evar_general}
	Let \(\w\in\mathbb{R}^N\) denote a portfolio weight vector for \(N\) assets and define the scalar portfolio return by \(X_{\w,t}=\wt\boldsymbol{X}_t\). Applying the L\'evy--Khintchine formula in Eq.~(\ref{charac_ftn_levy}) to Eq.~(\ref{mgf}) gives the moment-generating function of \(\boldsymbol{X}_t\) for \(\boldsymbol{u}\in\mathrm{dom}(M_{\boldsymbol{X}_t})\):
	\begin{align}
	\label{mgf_levy}
		M_{\boldsymbol{X}_t}(\boldsymbol{u}) = \exp\big(\Psi_{\boldsymbol{X}_t}(\boldsymbol{u})\big),
	\end{align}
	where the cumulant function \(\Psi_{\boldsymbol{X}_t}(\boldsymbol{u})\) is given by
	\begin{align}
	\label{cumulant_levy}
		\Psi_{\boldsymbol{X}_t}(\boldsymbol{u}) = t\Big[\frac{1}{2} \boldsymbol{u}^{\top}\boldsymbol{\Sigma}\boldsymbol{u} + \boldsymbol{\mu}^{\top}\boldsymbol{u} + \int_{\mathbb{R}^N} ({\rm e}^{\boldsymbol{u}^{\top}\boldsymbol{x}} - 1 - \boldsymbol{u}^{\top}\boldsymbol{x} \mathbbm{1}_{\|\boldsymbol{x}\| \leq 1}) \nu(d\boldsymbol{x})\Big].
	\end{align}
	
	With \(\boldsymbol{u}=u\w\) in Eq.~(\ref{mgf_levy}), the portfolio return moment-generating function can be written as
	\begin{align}
		M_{\wt\boldsymbol{X}_t}(u)=M_{\boldsymbol{X}_t}(u\w)=\exp(\Psi_{\boldsymbol{X}_t}(u\w)). 
	\end{align}
	The portfolio weights enter the cumulant function through the projected vector \(u\w\), which is the argument of \(\Psi_{\boldsymbol{X}_t}\).
	
	Substituting \(M_{\wt\boldsymbol{X}_t}(-u)=\exp(\Psi_{\boldsymbol{X}_t}(-u\w))\) into Eq.~(\ref{eq_evar}) gives the portfolio EVaR:
	\begin{align}
	\label{evar_levy}
		\text{EVaR}_{1-\eta}(\wt\boldsymbol{X}_t) = \inf_{u \in \mathcal{U}(\w)} \left\{\frac{\Psi_{\boldsymbol{X}_t}(-u\w) - \ln \eta}{u} \right\},
	\end{align}
	where the admissible scalar domain is given by
	\begin{align}
		\mathcal{U}(\w)=\{u>0:-u\w\in\mathrm{dom}(M_{\boldsymbol{X}_t})\}.
	\end{align}
	The domain \(\mathcal{U}(\w)\) changes with the portfolio weights because \(M_{\boldsymbol{X}_t}(-u\w)\) must be finite. Once \(\Psi_{\boldsymbol{X}_t}\) is specified, Eq.~(\ref{evar_levy}) is a one-dimensional minimization problem in \(u\) for each fixed \(\w\).
	
	When \(N=1\) and \(\w=1\), Eqs.~(\ref{cumulant_levy})--(\ref{evar_levy}) reduce to the univariate counterparts:
	\begin{align}
	\label{cumulant_levy_uni}
		\Psi_{X_t}(u) &= t\Big[ \frac{\sigma^2}{2}u^2 + \mu u + \int_{-\infty}^{\infty} ({\rm e}^{ux} - 1 - ux \mathbbm{1}_{|x| \leq 1}) \nu(dx)\Big],\\
	\label{evar_levy_uni}
		\text{EVaR}_{1-\eta}(X_t) &= \inf_{u \in \mathcal{U}} \left\{\frac{\Psi_{X_t}(-u) - \ln\eta}{u}\right\},
	\end{align}
	where \(\sigma^2\) is the scalar variance of the diffusion component. The corresponding EVaR objective function \(f\) is defined as
	\begin{align}
	\label{evar_obj_fn_levy}
		f(u)=\frac{\Psi_{X_t}(-u) - \ln\eta}{u},
	\end{align}
	whose infimum over \(\mathcal{U}\) gives the EVaR.

\subsection{Tempered stable processes}
\label{sec_evar_ts}
	Table~\ref{tab_cumulants} gives the cumulant functions and EVaR domains for the univariate Normal, CTS, and NTS processes. The CTS and NTS cumulants are obtained by substituting the L\'evy measures in Eqs.~(\ref{levy_measure_cts}) and (\ref{levy_measure_nts}) into Eq.~(\ref{cumulant_levy_uni}) (\cite{rachev2011financial}). The MNTS case is given in Section~\ref{sec_mnts_route}.
	
	\begin{table}[h]
	\centering
	\renewcommand{\arraystretch}{2.4}
	\caption{Cumulant function \(\Psi_{X_t}(u)\) and EVaR domain \(\mathcal{U}\) for the univariate Normal, CTS, and NTS processes. }
	\label{tab_cumulants}
	\begin{tabular}{p{2.4cm} p{8cm} p{4.6cm}}
		\hline
		Process & \(\Psi_{X_t}(u)\) & \(\mathcal{U}\) \\
		\hline
		Normal & \(t\big(\dfrac{\sigma^2 u^2}{2} + \mu u\big)\) & \((0,\infty)\) \\
		CTS & \(t\Big\{\big[\mu - C\Gamma(1-\alpha)(\lambda_+^{\alpha-1} - \lambda_-^{\alpha-1})\big] u \) \newline\hspace*{1em}\(+ C\Gamma(-\alpha)\big[(\lambda_+ - u)^\alpha - \lambda_+^\alpha + (\lambda_- + u)^\alpha - \lambda_-^\alpha\big]\Big\}\) & \((0,\lambda_-]\) \\
		NTS & \(t\Big\{(\mu - \beta) u - \dfrac{2 \theta}{\alpha}\big[(1 - \dfrac{\beta}{\theta} u - \dfrac{\gamma^2}{2\theta} u^2)^{\alpha/2} - 1\big]\Big\}\) & \(\Big(0,\dfrac{\beta+\sqrt{\beta^2+2\theta\gamma^2}}{\gamma^2}\Big]\) \\
		\hline
	\end{tabular}
	\end{table}
	
	The univariate EVaR is obtained by evaluating each cumulant in Table~\ref{tab_cumulants} at \(-u\) in Eq.~(\ref{evar_levy_uni}). The Normal case has a closed-form solution. Using the Normal cumulant of Table~\ref{tab_cumulants} in Eq.~(\ref{evar_levy_uni}), the EVaR expression for the Normal case is given by
	\begin{align}
	\label{evar_normal_prc}
		\text{EVaR}_{1-\eta}(X_{t}^{\text{Normal}}) = \inf_{u \in \mathcal{U}} \Big\{ t\big[ \tfrac{\sigma^2 u}{2} - \mu \big] -  u^{-1}\ln \eta \Big\}.
	\end{align}
	Since the objective is convex in \(u\) for \(0<\eta<1\), the first-order condition gives
	\begin{align}
	\label{evar_normal_inf_pt}
		u^*=\frac{\sqrt{-2\ln\eta/t}}{\sigma}.
	\end{align}
	Substituting \(u^*\) into the objective gives EVaR:
	\begin{align}
	\label{evar_normal_closed}
		\text{EVaR}_{1-\eta}(X^{\text{Normal}}_{t}) = -t\mu+\sigma\sqrt{-2t\ln\eta}.
	\end{align}
	When \(t=1\), \(\text{EVaR}_{1-\eta}(X^{\text{Normal}}) = -\mu+\sigma\sqrt{-2\ln \eta}\). Under the return--loss sign convention used in this paper, this expression agrees with the Normal EVaR in \cite{ahmadi2012entropic}.

\subsection{Numerical evaluation of the inner problem}
\label{sec_evar_numerical}
	For CTS and NTS processes, the first-order condition does not generally give a closed-form EVaR solution because of the non-integer powers of \(\alpha\). EVaR is evaluated numerically over the finite moment-generating-function domain of the fitted return model. For a fixed portfolio weight vector \(\w\), write
	\begin{align}
    		K_{\w}(u)=\Psi_{\boldsymbol{X}_t}(-u\w), \qquad f_{\w}(u)=\frac{K_{\w}(u)-\ln\eta}{u}, \qquad u\in\mathcal{U}(\w).
	\end{align}
	For the scalar EVaR problem, convexity of the cumulant-generating function implies that \(f_{\w}\) has at most one interior minimizer on its admissible domain (\cite{nedeltchev2026measuring}). If there is no interior minimum, the optimum may occur at a finite admissible endpoint. The numerical procedure searches the interior and also evaluates the endpoint when it is finite.

	Algorithm~\ref{alg_evar_eval} summarizes the numerical procedure. The interior search uses \([\epsilon,u_{\max}-\epsilon]\), where \(\epsilon>0\) avoids the singularity at \(u=0\) and direct cumulant evaluation near a finite boundary. The endpoint value is evaluated separately when it is finite. The interior interval is divided into \(P\) subintervals, and Brent's bounded minimization is applied to each.

\begin{algorithm}[H]
	\caption{Numerical evaluation of EVaR for tempered stable processes}
	\label{alg_evar_eval}
	\begin{algorithmic}[1]
	    \Require Family parameters \(\boldsymbol{\xi}\), confidence level \(\eta \in (0,1)\), time horizon \(t\), partition size \(P\), tolerance \(\epsilon>0\)
	    \State Determine the admissible upper bound \(u_{\max}\) from the domain of the moment-generating function associated with \(\boldsymbol{\xi}\)
	    \State Set the truncated interior search domain \(\mathcal{U}_{\epsilon}=[\epsilon,u_{\max}-\epsilon]\)
	    \State Define the cumulant function \(\Psi_{X_t}(u;\boldsymbol{\xi})\) for the selected process
	    \State Define the EVaR objective
	    \[
	        f(u)=\frac{\Psi_{X_t}(-u;\boldsymbol{\xi})-\ln\eta}{u},
	        \qquad u\in\mathcal{U}_{\epsilon}.
	    \]
	    \State Partition \(\mathcal{U}_{\epsilon}\) into \(P\) subintervals \([u_0,u_1], [u_1,u_2], \ldots, [u_{P-1},u_P]\)
	    \For{\(p=1,\ldots,P\)}
	        \State Apply Brent's bounded minimization on \([u_{p-1},u_p]\) to obtain \(u_p^*\)
	        \State Compute \(f_p=f(u_p^*)\)
	    \EndFor
	    \State Set \(f_{\mathrm{bdry}}\gets+\infty\)
	    \If{the analytic value at the admissible finite endpoint is finite}
	        \State \(f_{\mathrm{bdry}}\gets\{\Psi_{X_t}(-u_{\max};\boldsymbol{\xi})-\ln\eta\}/u_{\max}\)
	    \EndIf
	    \State \Return \(\min\{f_1,\ldots,f_P,f_{\mathrm{bdry}}\}\)
	\end{algorithmic}
\end{algorithm}

	The lower truncation does not exclude an optimizer near zero because \(K_{\w}(u)=O(u)\) as \(u\downarrow0\) while \(-\ln\eta>0\), so that \(f_{\w}(u)\to+\infty\). Since the scalar objective is unimodal on its admissible interior, the \(P\) subintervals are used to reduce numerical sensitivity, not to search for multiple local minima.

\section{EVaR portfolio optimization}
\label{sec_ica_evar}
	This section presents two EVaR portfolio optimization approaches based on MNTS projection and ICA factorization. It also defines the entropic reward--risk objectives used in the portfolio problems. 

\subsection{Computational motivation}
\label{sec_setup}
	A direct parametric implementation constructs the return series for each candidate weight vector, fits the assumed univariate portfolio distribution, and evaluates EVaR from the fitted parameters. Repeating this procedure over the feasible weight set is computationally intensive because the fitted distributional parameters and admissible moment-generating-function domain generally change with the portfolio weights.
	
	The two approaches below avoid repeated portfolio-level fitting. The MNTS approach uses closure under linear projection, and the ICA approach builds the portfolio cumulant from fitted component-specific marginal distributions. In both cases, parameters estimated from the in-sample asset or component returns are reused across candidate weights to construct the corresponding portfolio cumulant and admissible moment-generating-function domain.
	
\subsection{MNTS-based approach}
\label{sec_mnts_route}
	The MNTS model has previously been used for portfolio optimization and risk allocation. In particular, \cite{kim2022portfolio} exploits the MNTS structure to characterize the distribution of portfolio returns, develops a portfolio optimization framework based on dispersion and asymmetric tail risk, and derives marginal VaR and marginal CVaR for risk budgeting. In this paper, the projected univariate NTS parameters are used directly in the portfolio cumulant-generating function and its admissible moment-generating-function domain. These two quantities determine the portfolio EVaR.
	
	The MNTS process introduced in Section~\ref{sec_mnts_dist} is closed under linear projection. Its single shared subordinator implies that any linear projection \(\wt\boldsymbol{X}^{\text{MNTS}}_t\) is a univariate NTS process. From Eq.~(\ref{mnts_def}), the linear projection is given by
	\begin{align}
	\label{mnts_projection}
	\wt \boldsymbol{X}^{\text{MNTS}}_t = (\wt\boldsymbol{\mu}) t + (\wt\boldsymbol{\beta})(\mathcal{T}_t-t)+\wt\mathrm{diag}(\boldsymbol{\gamma})\boldsymbol{B}_{\mathcal{T}_t},
	\end{align}
	where the Brownian term \(\wt\mathrm{diag}(\boldsymbol{\gamma})\boldsymbol{B}_{\mathcal{T}_t}\) conditional on \(\mathcal{T}_t\) is normally distributed with mean zero and variance \(\mathcal{T}_t\wt\boldsymbol{\Sigma}\w\) for \(\boldsymbol{\Sigma}=\mathrm{diag}(\boldsymbol{\gamma})\boldsymbol{\rho}\,\mathrm{diag}(\boldsymbol{\gamma})\). Comparing Eq.~(\ref{mnts_projection}) with the univariate NTS construction in Section~\ref{sec_prelim}, the projected variable \(\wt\boldsymbol{X}^{\text{MNTS}}_t\) follows the univariate NTS process \(\mathrm{NTS}(\bar\alpha,\bar\theta,\bar\beta,\bar\gamma,\bar\mu)\) with the projected parameters
	\begin{align}
	\label{mnts_projected_params}
		\bar\alpha = \alpha,\quad \bar\theta = \theta, \quad \bar\beta = \wt \boldsymbol{\beta}, \quad \bar\gamma = \sqrt{\wt\boldsymbol{\Sigma}\w}, \quad \bar\mu = \wt \boldsymbol{\mu}.
	\end{align}
	
	For \(\bar\gamma^2>0\), the weight-dependent EVaR domain follows directly from the NTS domain in Table~\ref{tab_cumulants}:
	\begin{align}
	\label{mnts_domain_explicit}
		\mathcal{U}_{\mathrm{MNTS}}(\w)&=\left(0,\,u_{\max}^{\mathrm{MNTS}}(\w)\right],\\
		u_{\max}^{\mathrm{MNTS}}(\w)&=\frac{\bar\beta+\sqrt{\bar\beta^2+2\theta\bar\gamma^2}}{\bar\gamma^2}.
	\end{align}
	Both the projected cumulant and its admissible endpoint change explicitly with \(\w\) via \(\bar\beta\) and \(\bar\gamma^2\). 

	Using the projected parameters of Eq.~(\ref{mnts_projected_params}) in the NTS row of Table~\ref{tab_cumulants} and the L\'evy EVaR formula in Eq.~(\ref{evar_levy_uni}), the MNTS portfolio EVaR is given by
	\begin{align}
	\label{evar_mnts_prc}
		\mathrm{EVaR}^{\mathrm{MNTS}}_{1-\eta}(\wt\boldsymbol{X}^{\mathrm{MNTS}}_t)=\inf_{u\in\mathcal{U}_{\mathrm{MNTS}}(\w)}\left\{\frac{\Psi^{NTS}_{X_t}(-u;\bar\alpha,\bar\theta,\bar\beta,\bar\gamma,\bar\mu)-\ln\eta}{u}\right\},
	\end{align}
	where \(\Psi^{NTS}_{X_t}\) is the NTS cumulant in Table~\ref{tab_cumulants} evaluated at the projected parameters, and \(\mathcal{U}_{\mathrm{MNTS}}(\w)\) is the corresponding projected NTS domain. When \(N=1\) and \(\w=1\), Eq.~(\ref{evar_mnts_prc}) reduces to the univariate NTS EVaR. Once the MNTS parameters \((\alpha,\theta,\boldsymbol{\beta},\boldsymbol{\gamma},\boldsymbol{\mu},\boldsymbol{\rho})\) are estimated from the joint return data, the EVaR for any portfolio weight can be computed without re-estimating distributional parameters across weights. The inner one-dimensional infimum in Eq.~(\ref{evar_mnts_prc}) is evaluated using Algorithm~\ref{alg_evar_eval} with the projected NTS parameters and the corresponding admissible domain.

\subsection{ICA-based approach}
\label{sec_ica_route}
	The ICA-based approach represents the joint return distribution through component-specific marginals modeled by L\'evy, CTS, or NTS processes. We apply the FastICA algorithm (\cite{hyvarinen2000independent}) to the centered in-sample return matrix \(\mathbf{X}-\boldsymbol{m}\mathbf{1}_{T}^{\top}\) with the sample mean vector \(\boldsymbol{m}\in\mathbb{R}^{N}\). The transformation to the component matrix \(\mathbf{S}\) is given by
	\begin{align}
		\label{eq_ica_unmixing}
		\mathbf{S} = \mathbf{G}\left(\mathbf{X}-\boldsymbol{m}\mathbf{1}_{T}^{\top}\right),
	\end{align}
	where \(\mathbf{S}\) denotes the \(J \times T\) matrix of estimated independent components and \(\mathbf{G}\) is the \(J \times N\) matrix, also referred to as the unmixing matrix. The unmixing matrix is estimated to maximize statistical independence among the components.
	
	The observed return matrix can be reconstructed from the independent components as
	\begin{align}
		\label{eq_ica_mixing}
		\mathbf{X} = \boldsymbol{m}\mathbf{1}_{T}^{\top} + \mathbf{A}\mathbf{S},
	\end{align}
	where \(\mathbf{A}\) is the \(N \times J\) matrix, referred to as the mixing matrix. The mixing matrix characterizes how the independent components are combined to reconstruct the centered dependent asset returns, and the location vector \(\boldsymbol{m}\) captures the one-period mean not absorbed into the independent components. For the L\'evy-process extension used below, we write \(\boldsymbol{X}_t=t\boldsymbol{m}+\mathbf{A}\boldsymbol{S}_t\).

	Under the ICA construction, the components are modeled as mutually independent, and the scalar portfolio return \(X_{\w,t}=t\wt\boldsymbol{m}+\sum_{i=1}^{J}(\wt\mathbf{A})_iS_t^{(i)}\) has moment-generating function:
	\begin{align}
		M_{X_{\w,t}}(u)= e^{ut\wt\boldsymbol{m}}\,\prod_{i=1}^{J} M_{S_t^{(i)}}\big(u (\wt \mathbf{A})_i\big).
	\end{align}
	The portfolio cumulant is the affine location term \(ut\wt\boldsymbol{m}\) plus the sum of the independent component cumulants.
	
	The corresponding portfolio EVaR is given by
	\begin{align}
	\label{eq_evar_ica}
		\textrm{EVaR}_{1-\eta}(X_{\w,t}) = -t\wt\boldsymbol{m}+\inf_{u\in \mathcal{U}_{\mathrm{ICA}}(\w)} \left\{ \frac{1}{u} \bigg(\sum_{i=1}^{J}\ln M_{S_t^{(i)}}(-u(\wt \mathbf{A})_i)- \ln{\eta}\bigg) \right\}.
	\end{align}
	FastICA produces the mixing and component matrices used to fit the component-level distributions.
	
	Using \(\tw=\mathbf{A}^{\top}\w\), the ICA-based portfolio EVaR can then be written as
	\begin{align}
	\label{evar_ica_route}
	\mathrm{EVaR}^{\mathrm{ICA}}_{1-\eta}(\wt\boldsymbol{X}_t)=
	-t\wt\boldsymbol{m}+\inf_{u\in\mathcal{U}_{\mathrm{ICA}}(\w)}
	\left\{
	\frac{
	\sum_{i=1}^{J}\Psi_{S^{(i)}_t}(-u\tilde{w}_i)
	-\ln\eta
	}{u}
	\right\},
	\end{align}
	where  \(\tilde{w}_i\) is the \(i\)-th component of \(\tw\) and the admissible scalar domain is given by
	\begin{align}
	\label{ica_domain}
		\mathcal{U}_{\mathrm{ICA}}(\w)=\bigcap_{i=1}^{J} \left\{u>0: -u\tilde{w}_i\in\mathrm{dom}(M_{S^{(i)}_t})\right\}.
	\end{align}
	For the fitted CTS and NTS component families, the intersection has an explicit endpoint. For CTS processes, define
	\begin{align}
	\label{ica_cts_domain_endpoint}
		u^{\mathrm{CTS}}_{i,\max}(\tilde w_i)=
		\begin{cases}
		\lambda_{i,-}/\tilde w_i, & \tilde w_i>0,\\
		\lambda_{i,+}/|\tilde w_i|, & \tilde w_i<0,\\
		+\infty, & \tilde w_i=0,
		\end{cases}
	\end{align}
	and for NTS processes, define
	\begin{align}
	\label{ica_nts_domain_endpoint}
		u^{\mathrm{NTS}}_{i,\max}(\tilde w_i)=
		\begin{cases}
		\dfrac{\beta_i+\sqrt{\beta_i^2+2\theta_i\gamma_i^2}}{\gamma_i^2\tilde w_i}, & \tilde w_i>0,\\
		\dfrac{-\beta_i+\sqrt{\beta_i^2+2\theta_i\gamma_i^2}}{\gamma_i^2|\tilde w_i|}, & \tilde w_i<0,\\
		+\infty, & \tilde w_i=0.
		\end{cases}
	\end{align}
	The EVaR admissible domain for a CTS or NTS component specification can be written as
	\begin{align}
	\label{ica_domain_explicit}
		\mathcal{U}_{\mathrm{ICA}}(\w)=\left(0,\,\min_{1\le i\le J}u_{i,\max}(\tilde w_i)\right],
	\end{align}
	where \(u_{i,\max}\) is chosen from Eq.~(\ref{ica_cts_domain_endpoint}) or Eq.~(\ref{ica_nts_domain_endpoint}) according to the component family. The ICA admissible endpoint varies with the portfolio weights through the projected component exposures \(\tw=\mathbf{A}^{\top}\w\).

	For CTS and NTS components, \(\Psi_{S^{(i)}_t}\) is specified by the corresponding row of Table~\ref{tab_cumulants} with component-specific parameters. The inner one-dimensional infimum in Eq.~(\ref{evar_ica_route}) is evaluated using Algorithm~\ref{alg_evar_eval} with the aggregated cumulant and the intersection of the per-component domains. Once the ICA mixing matrix and the component-level distributional parameters are estimated from the in-sample data, the portfolio EVaR can be evaluated across candidate weights without reconstructing the portfolio return series or re-estimating the portfolio-level distribution.

	For example, suppose that the independent components follow L\'evy processes with L\'evy triplets of \((\sigma_i^2, \nu_i, \mu_i)\) for \(i=1,\cdots,J\). The portfolio EVaR under L\'evy-driven components can be written as
\begin{align}
	\label{evar_levy_ica}
		&\text{EVaR}_{1-\eta}(X^{\text{L\'evy}}_{\w,t}) \nonumber\\
		&= -t\wt(\boldsymbol{m} + \mathbf{A}\boldsymbol{\mu}_{S}) + \inf_{u \in \mathcal{U}_{\mathrm{ICA}}(\w)} \left\{ t\sum_{i=1}^{J}\Big[ \frac{(\sigma_i(\wt \mathbf{A})_i)^2}{2}u + \int_{-\infty}^{\infty} (\frac{{\rm e}^{-u(\wt \mathbf{A})_i x} - 1}{u} + (\wt \mathbf{A})_i x \mathbbm{1}_{|x| \leq 1}) \nu_i(dx) \Big] - \frac{1}{u}\ln \eta \right\}\nonumber\\
		&= -t\wt\boldsymbol{m} - t \twt\boldsymbol{\mu}_{S} + \inf_{u \in \mathcal{U}_{\mathrm{ICA}}(\w)} \left\{ t\sum_{i=1}^{J}\Big[ \frac{(\sigma_i\tilde{w}_i)^2}{2}u + \int_{-\infty}^{\infty} (\frac{{\rm e}^{-u \tilde{w}_i x} - 1}{u} + \tilde{w}_i x \mathbbm{1}_{|x| \leq 1}) \nu_i(dx) \Big] -  \frac{1}{u}\ln \eta \right\},
	\end{align}
	where \(\boldsymbol{\mu}_{S}=(\mu_1,\ldots,\mu_J)^{\top}\) for the component drift vector.
	
	For independent components following CTS processes with the triplets of \((0, \nu_i, \mu_i)\), where \(\nu_i\) is the L\'evy measure associated with \(\mathrm{CTS}(\alpha_i, C_i, \lambda_{i,+}, \lambda_{i,-}, \mu_i)\) for \(i=1,\cdots, J\), the portfolio EVaR is given by
\begin{align}
	\label{evar_cts_ica}
		&\textrm{EVaR}_{1-\eta}(X^{\text{CTS}}_{\w,t}) \nonumber \\
		&= -t\wt(\boldsymbol{m} + \mathbf{A} \boldsymbol{\tilde{\mu}}_{S}) + \inf_{u\in \mathcal{U}_{\mathrm{ICA}}(\w)} \bigg\{\sum_{i=1}^{J}\Big(\frac{t C_i \Gamma(-\alpha_i)}{u} \big[(\lambda_{i,+} + u(\wt \mathbf{A})_i)^{\alpha_i} - \lambda_{i,+}^{\alpha_i} + (\lambda_{i,-} - u(\wt \mathbf{A})_i)^{\alpha_i} - \lambda_{i,-}^{\alpha_i}\big]\Big)-\frac{1}{u} \ln{\eta}\bigg\} \nonumber \\
		&= -t\wt\boldsymbol{m} - t \twt \boldsymbol{\tilde{\mu}}_{S}+ \inf_{u\in \mathcal{U}_{\mathrm{ICA}}(\w)} \bigg\{\sum_{i=1}^{J}\Big(\frac{t C_i \Gamma(-\alpha_i)}{u} \big[(\lambda_{i,+} + u\tilde{w}_i)^{\alpha_i} - \lambda_{i,+}^{\alpha_i} + (\lambda_{i,-} - u \tilde{w}_i)^{\alpha_i} - \lambda_{i,-}^{\alpha_i}\big]\Big)-\frac{1}{u} \ln{\eta}\bigg\},
	\end{align}
	where \(\boldsymbol{\tilde{\mu}}_{S}=(\tilde\mu_1,\ldots,\tilde\mu_J)^{\top}\) is the component adjusted-drift vector with
	\begin{align}
		\tilde{\mu}_i=\mu_i - C_i \Gamma(1-\alpha_i)(\lambda_{i,+}^{\alpha_i-1} - \lambda_{i,-}^{\alpha_i-1}).
	\end{align}
	
	For independent components following NTS processes with L\'evy triplets \((0, \nu_i, \mu_i)\), where \(\nu_i\) is the L\'evy measure associated with \(\mathrm{NTS}(\alpha_i, \theta_i, \beta_i, \gamma_i, \mu_i)\) for \(i=1,\cdots, J\), the portfolio EVaR expression follows by combining the NTS cumulant in Table~\ref{tab_cumulants} with Eq.~(\ref{eq_evar_ica}):
\begin{align}
	\label{evar_nts_ica}
		&\textrm{EVaR}_{1-\eta}(X^{\text{NTS}}_{\w,t}) \nonumber \\
		&= -t\wt(\boldsymbol{m} + \mathbf{A} (\boldsymbol{\mu}_{S} -\boldsymbol{\beta}_{S})) - \sup_{u\in \mathcal{U}_{\mathrm{ICA}}(\w)} \bigg\{\sum_{i=1}^{J}\frac{2 t \theta_i}{u\alpha_i} \big[ (1 + \frac{\beta_i (\wt \mathbf{A})_i }{\theta_i} u - \frac{(\gamma_i (\wt \mathbf{A})_i)^2}{2\theta_i} u^2)^{\frac{\alpha_i}{2}} -1\big]+\frac{1}{u} \ln{\eta}\bigg\} \nonumber \\
		&= -t\wt\boldsymbol{m} -t \twt (\boldsymbol{\mu}_{S} -\boldsymbol{\beta}_{S}) - \sup_{u\in \mathcal{U}_{\mathrm{ICA}}(\w)} \bigg\{\sum_{i=1}^{J}\frac{2 t \theta_i}{u\alpha_i} \big[ (1 + \frac{\beta_i \tilde{w}_i}{\theta_i} u  - \frac{(\gamma_i\tilde{w}_i)^2}{2\theta_i} u^2)^{\frac{\alpha_i}{2}} -1\big]+\frac{1}{u} \ln{\eta}\bigg\},
	\end{align}
	where \(\boldsymbol{\mu}_{S}=(\mu_1,\ldots,\mu_J)^{\top}\) and \(\boldsymbol{\beta}_{S}=(\beta_1,\ldots,\beta_J)^{\top}\).

\subsection{Parameter estimation}
\label{sec_estimation}
	The MNTS parameters are estimated from the asset return series following the MNTS specification in \cite{kim2012measuring, kim2022portfolio}. A univariate NTS distribution is first fitted to each asset return series using the ECDF--MSE estimator. Since the MNTS model has a common tempered stable subordinator, the common parameters \((\alpha,\theta)\) are set to the cross-sectional averages of the marginal estimates. Conditional on these common parameters, the marginal skewness parameters are re-estimated. The correlation matrix of the Brownian components is then obtained from the sample covariance matrix by matching the second moments of the MNTS model. The fitted marginal parameters are converted back to the scale of the original returns and used with the estimated correlation matrix in the MNTS projection of Section~\ref{sec_mnts_route}.

	For the ICA specifications, the return series are first centered and FastICA is applied to estimate the mixing matrix and independent components (\cite{hyvarinen2000independent}). All \(N\) components are retained. An NTS or CTS distribution is then fitted independently to each component using the same ECDF--MSE estimator. The NTS and CTS parameterizations follow \cite{rachev2011financial}. The fitted component parameters are converted back to the component scale and used with the estimated mixing matrix and mean vector in the ICA portfolio EVaR calculation of Section~\ref{sec_ica_route}.

\subsection{EVaR-based reward--risk objectives}
\label{sec_entropic_ratios}
	In addition to minimum EVaR, the portfolio problems use E-STAR and E-Rachev ratios. These ratios are obtained from entropic extensions of the STAR and Rachev ratios in Table~\ref{tab_reward_risk} by replacing CVaR with EVaR. Table~\ref{tab_entropic_reward_risk} gives the corresponding reward and risk functionals.
	
\begin{table}[ht]
\centering
\caption{Entropic reward--risk objectives as ratios of reward and risk functionals.}
\label{tab_entropic_reward_risk}
\begin{tabular}{lll}
\hline
Objective & Reward functional \(\mathcal{G}(X)\) & Risk functional \(\mathcal{D}(X)\)  \\
\hline
E-STAR & \(\mathbb{E}[X]\) & \(\mathrm{EVaR}_{1-\eta}(X)\) \\
E-Rachev & \(\mathrm{EVaR}_{1-\zeta}(-X)\) & \(\mathrm{EVaR}_{1-\eta}(X)\) \\
\hline
\end{tabular}
\end{table}

	These entropic reward--risk measures are defined as
	\begin{align}
	\label{eq_estar}
		\mathrm{E\text{-}STAR}_{1-\eta}(X_t)&=\frac{\mathbb{E}[X_t]}{\mathrm{EVaR}_{1-\eta}(X_t)},\\
	\label{eq_erachev}
		\mathrm{E\text{-}Rachev}_{1-\zeta,1-\eta}(X_t)	&=	\frac{\mathrm{EVaR}_{1-\zeta}(-X_t)}	{\mathrm{EVaR}_{1-\eta}(X_t)}.
	\end{align}
	The E-STAR objective function measures expected return per unit of entropic downside risk. The E-Rachev objective function compares entropic upper-tail reward with entropic downside-tail risk. These two objectives and the minimum-EVaR optimization are used below.

\begin{rmk}[Normal benchmark equivalence]
	\label{cor_normal_ratio_equivalence}
	The Normal specification gives an analytical benchmark for the reward--risk objectives. Let \(X_{\w}=\w^{\top}\boldsymbol{X}\sim\mathcal{N}(\mu_{\w},\sigma_{\w}^2)\) with \(\sigma_{\w}>0\), define \(s_{\w}=\mu_{\w}/\sigma_{\w}\) and \(k_a=\sqrt{-2\ln a}\), and suppose that \(k_\eta-s_{\w}>0\) on the feasible set. The Normal EVaR formula implies
	\begin{align}
		\mathrm{E\text{-}STAR}_{1-\eta}(X_{\w})&=\frac{s_{\w}}{k_\eta-s_{\w}},	\\
		\mathrm{E\text{-}Rachev}_{1-\zeta,1-\eta}(X_{\w})	&=\frac{k_\zeta+s_{\w}}{k_\eta-s_{\w}}.
	\end{align}
	Since their derivatives with respect to \(s\) are \(k_\eta/(k_\eta-s)^2\) and \((k_\eta+k_\zeta)/(k_\eta-s)^2\), respectively, both objectives are strictly increasing functions of \(s_{\w}\). When the maximum-Sharpe objective uses the same mean-return convention and feasible set, the Normal E-STAR, Normal E-Rachev, and maximum-Sharpe portfolios have the same set of optimal weights. We use this identity below as an analytical benchmark. For Gaussian minimum-EVaR, \cite{luxenberg2024portfolio} gives the related Markowitz-type mean--standard-deviation representation.
\end{rmk}

\subsection{Optimization formulations}
\label{sec_opt_objectives}

	Let \(\mathcal{A}\in\{\mathrm{MNTS},\mathrm{ICA}\}\) denote the EVaR evaluation approach. We write \(\mathrm{EVaR}^{\mathcal{A}}_{1-\eta}(\wt\boldsymbol{X})\) for the portfolio EVaR computed under approach \(\mathcal{A}\). For the MNTS approach, the EVaR is evaluated using the projected NTS parameters in Eq.~(\ref{mnts_projected_params}) and the corresponding formula in Eq.~(\ref{evar_mnts_prc}). For the ICA approach, the EVaR is evaluated using the component cumulants and projected component weights \(\tw=\mathbf{A}^{\top}\w\) in Eq.~(\ref{evar_ica_route}).

	To express the E-STAR and E-Rachev objectives defined in Section~\ref{sec_entropic_ratios} as functions of the portfolio weights under each EVaR evaluation approach, define
	\begin{align}
		\mathcal{G}_{\mathrm{ES}}(\w)	&=\wt\mathbb{E}[\boldsymbol{X}],\label{eq_reward_es_w}\\
		\mathcal{D}_{\eta}^{\mathcal{A}}(\w)&=\mathrm{EVaR}^{\mathcal{A}}_{1-\eta}(\wt\boldsymbol{X}),\label{eq_downside_evar_w}\\
		\mathcal{G}_{\mathrm{ER},\zeta}^{\mathcal{A}}(\w)&=\mathrm{EVaR}^{\mathcal{A}}_{1-\zeta}(-\wt\boldsymbol{X}).
		\label{eq_reward_er_w}
	\end{align}
	Here \(\mathcal{G}_{\mathrm{ES}}\) is the reward term of the E-STAR ratio, \(\mathcal{D}_{\eta}^{\mathcal{A}}\) is the common downside-risk term of the E-STAR and E-Rachev ratios, and \(\mathcal{G}_{\mathrm{ER},\zeta}^{\mathcal{A}}\) is the upper-tail reward term of the E-Rachev ratio. The expected-return functional \(\mathcal{G}_{\mathrm{ES}}\) is independent of the multivariate representation. The two EVaR-based functionals depend on whether EVaR is evaluated under the MNTS or ICA construction.

	The feasible set of portfolio weights is given by
	\begin{align}
		\mathcal{W} = \left\{\w \in \mathbb{R}^N : \wt\boldsymbol{e} = 1,\; 	\wt\mathbb{E}[\boldsymbol{X}] \ge R_*,\; 	\w \in \mathcal{C}\right\},
	\end{align}
	where \(\boldsymbol{e}=(1,1,\cdots,1)\), \(R_*\) is the target expected return on the portfolio, and \(\mathcal{C}\) collects additional investor-specific constraints such as long-only, box, or leverage constraints.

	The minimum-EVaR portfolio under approach \(\mathcal{A}\) is defined as
	\begin{align}
		\w^{*,\mathcal{A}}_{\mathrm{ME}} = \arg\min_{\w \in \mathcal{W}} \mathcal{D}_{\eta}^{\mathcal{A}}(\w).
	\end{align}

	Convexity of minimum-EVaR portfolio optimization is established in prior work, including the sample-based formulation of \cite{ahmadi2019portfolio}. For Gaussian mixture returns, \cite{luxenberg2024portfolio} obtains a convex minimum-EVaR program through the reciprocal change of variable and the perspective of the cumulant-generating function. The same perspective argument applies directly to the cumulant representation used by both parametric constructions in this paper. Introducing \(v=1/u\) gives
	\begin{align}
	\label{eq_persp}
		F(\w,v)=v\,\Psi_{\boldsymbol{X}_t}(-\w/v)+v\ln(1/\eta),
		\qquad
		\mathrm{dom}\,F=\{(\w,v):v>0,\;-\w/v\in\mathrm{dom}(M_{\boldsymbol{X}_t})\}.
	\end{align}
	Since the cumulant-generating function is convex on its domain, its perspective is jointly convex in \((\w,v)\) on \(\mathrm{dom}\,F\), and partial minimization over the admissible \(v\)-section preserves convexity in \(\w\). The relation \(F(c\w,cv)=cF(\w,v)\) for \(c>0\) also implies that the resulting downside-risk functional \(\mathcal{D}_{\eta}^{\mathcal{A}}\) is positively homogeneous. Minimum-EVaR optimization over a convex feasible set is a convex program under both the MNTS and ICA approaches and may be solved either by the inner--outer decomposition used here or as a joint convex problem in \((\w,v)\). When \(\mathcal{C}\) is convex, any numerically converged minimum-EVaR solution satisfying the optimality conditions is globally optimal.

	The E-STAR portfolio maximizes the expected return per unit of entropic downside risk:
	\begin{align}
		\w^{*,\mathcal{A}}_{\mathrm{ES}} = 
		\arg\max_{\w \in \mathcal{W}} \mathrm{E\text{-}STAR}^{\mathcal{A}}_{1-\eta}(\wt\boldsymbol{X}),
	\end{align}
	where the corresponding objective is given by
	\begin{align}
		\mathrm{E\text{-}STAR}^{\mathcal{A}}_{1-\eta}(\wt\boldsymbol{X})
		=
		\frac{\mathcal{G}_{\mathrm{ES}}(\w)}{\mathcal{D}_{\eta}^{\mathcal{A}}(\w)}.
	\end{align}
	Because \(\mathcal{G}_{\mathrm{ES}}\) is linear and \(\mathcal{D}_{\eta}^{\mathcal{A}}\) is positive and convex on the feasible set, the E-STAR superlevel sets are convex for nonnegative ratio levels. We solve the resulting fractional program by Dinkelbach iterations (\cite{dinkelbach1967nonlinear}). For nonnegative ratio parameters, each parametric subproblem is a convex optimization problem. Under compactness of \(\mathcal{W}\), positivity of \(\mathcal{D}_{\eta}^{\mathcal{A}}\) on \(\mathcal{W}\), and a nonnegative optimal ratio, the standard Dinkelbach argument ensures convergence to the global fractional optimum.

	The E-Rachev portfolio maximizes the ratio of entropic upper-tail reward to entropic downside-tail risk:
	\begin{align}
		\w^{*,\mathcal{A}}_{\mathrm{ER}} = 
		\arg\max_{\w \in \mathcal{W}} \mathrm{E\text{-}Rachev}^{\mathcal{A}}_{1-\zeta,1-\eta}(\wt\boldsymbol{X}),
	\end{align}
	where the corresponding objective can be expressed as
	\begin{align}
		\mathrm{E\text{-}Rachev}^{\mathcal{A}}_{1-\zeta,1-\eta}(\wt\boldsymbol{X})
		=
		\frac{\mathcal{G}_{\mathrm{ER},\zeta}^{\mathcal{A}}(\w)}
		{\mathcal{D}_{\eta}^{\mathcal{A}}(\w)}.
	\end{align}
	Unlike the E-STAR ratio, both \(\mathcal{G}_{\mathrm{ER},\zeta}^{\mathcal{A}}\) and \(\mathcal{D}_{\eta}^{\mathcal{A}}\) are convex functions of the portfolio weights, and their ratio is not generally quasiconcave. We solve the E-Rachev problem by minimizing the negative ratio from multiple feasible starting portfolios and retain the candidate with the largest attained E-Rachev value. Using multiple starting portfolios reduces dependence on a single starting point but does not guarantee a global optimum. The upper-tail EVaR does not require refitting the distribution to sign-inverted returns: sign inversion follows directly from the fitted tempered stable parameters, for example by interchanging the positive and negative tempering parameters for CTS processes and reversing the skewness parameter for NTS processes, with the corresponding location and skewness transformations applied in the multivariate representations. The same fitted model can therefore be used to evaluate both tail functionals.

	The optimization formulation is the same for the MNTS and ICA approaches. Only the EVaR evaluation differs between the two representations. Table~\ref{tab_objective_taxonomy} summarizes the three cases.

\begin{table}[!t]
\centering
\caption{Structure of the entropic portfolio objectives and the associated solution guarantees.}
\label{tab_objective_taxonomy}
\small
\begin{tabular}{llll}
\toprule
Objective & Structure in \(\w\) & Method & Guarantee \\
\midrule
Minimum EVaR & Convex & Convex minimization & Global optimum \\
E-STAR & Convex superlevel sets for \(q\ge0\) & Dinkelbach iterations & Global under stated conditions \\
E-Rachev & Not quasiconcave in general & Multi-start optimization & Local optima \\
\bottomrule
\end{tabular}
\end{table}
	
\subsection{CVaR benchmark portfolios}
\label{sec_cvar_baseline}
	The CVaR-based portfolios are included only as matched benchmarks for the EVaR portfolio strategies. Within each MNTS or ICA specification, they use the same fitted
distributional parameters, feasible set, estimation window, and rebalancing schedule as the corresponding entropic portfolio. For portfolio loss \(L=-\w^{\top}\boldsymbol{X}\), VaR is computed by one-dimensional Gil--Pelaez inversion of the fitted loss characteristic function, and CVaR is evaluated from the associated average-value-at-risk integral by FFT (\cite{rockafellar2000optimization}). The MNTS characteristic function follows from linear projection, and the ICA characteristic function is formed from the product of the fitted component characteristic functions.

	The optimal CVaR, STAR, and Rachev portfolios mirror the optimal EVaR, E-STAR, and E-Rachev objectives. Minimum-CVaR is solved by sequential least-squares programming with numerical differentiation, STAR by Dinkelbach iterations (\cite{dinkelbach1967nonlinear}), and Rachev by multi-start sequential least-squares programming. These matched portfolios allow direct EVaR--CVaR comparisons under the same fitted model and portfolio setup.

\section{Empirical results}
\label{sec_result}
	This section presents a rolling backtest of the EVaR portfolio methods using U.S. sector ETFs. The analysis compares matched EVaR and CVaR portfolios under Normal and tempered stable specifications, ICA and MNTS representations, transaction costs, market regimes, and portfolio turnover.

\subsection{Dataset and methodology}
\subsubsection{Dataset for backtesting}
\label{sec_dataset}
	The universe consists of the eleven State Street Select Sector SPDR ETFs, which partition large-capitalization U.S.\ equities into sector portfolios. Nine of the funds have traded since December~1998, and the real estate (XLRE) and communication services (XLC) funds were introduced in 2015 and 2018, respectively. Each fund enters the investable set after one full year of return history is available, so the universe expands from nine to eleven funds over the sample. We use the sector ETFs because the universe does not contain a low-volatility asset class in which a downside-risk minimizer can concentrate. The assets have broadly comparable volatility but different tail behavior. The funds are also liquid and directly investable, so turnover and transaction-cost comparisons are meaningful. Adjusted daily prices are sourced from Yahoo Finance beginning at each fund's inception. After the initial twelve-month estimation window, the out-of-sample period begins in January~2000 and runs through March~2026, approximately twenty-six years.

\subsubsection{Portfolio construction and strategy set}
\label{sec_portfolio_construction}
	We evaluate thirty-one portfolios spanning standard allocation benchmarks, Normal entropic benchmarks, and tempered stable strategies under MNTS projection, ICA\(+\)NTS, and ICA\(+\)CTS. Table~\ref{tab_portfolios} summarizes the resulting strategy set.

\begin{table}[ht]
\centering
\caption{Portfolio strategies in the empirical analysis}
\label{tab_portfolios}
\small
\begin{tabular}{lll}
\toprule
Group & Portfolio class & Strategies included \\
\midrule
Standard benchmarks
& Stand-alone allocation
& EW, MinVar, MaxSharpe \\

Normal benchmarks
& Thin-tailed entropic allocation
& \(\mathrm{EVaR}_{95}\), \(\mathrm{E\text{-}STAR}_{95}\), \(\mathrm{E\text{-}Rachev}_{95,95}\), \(\mathrm{E\text{-}Rachev}_{50,95}\) \\

\multirow[t]{2}{*}{MNTS projection}
& Main EVaR specifications
& \(\mathrm{EVaR}_{95}\), \(\mathrm{E\text{-}STAR}_{95}\), \(\mathrm{E\text{-}Rachev}_{95,95}\), \(\mathrm{E\text{-}Rachev}_{50,95}\) \\
& CVaR benchmarks
& \(\mathrm{CVaR}_{95}\), \(\mathrm{STAR}_{95}\), \(\mathrm{Rachev}_{95,95}\), \(\mathrm{Rachev}_{50,95}\) \\

\multirow[t]{2}{*}{ICA + NTS}
& Main EVaR specifications
& \(\mathrm{EVaR}_{95}\), \(\mathrm{E\text{-}STAR}_{95}\), \(\mathrm{E\text{-}Rachev}_{95,95}\), \(\mathrm{E\text{-}Rachev}_{50,95}\) \\
& CVaR benchmarks
& \(\mathrm{CVaR}_{95}\), \(\mathrm{STAR}_{95}\), \(\mathrm{Rachev}_{95,95}\), \(\mathrm{Rachev}_{50,95}\) \\

\multirow[t]{2}{*}{ICA + CTS}
& Main EVaR specifications
& \(\mathrm{EVaR}_{95}\), \(\mathrm{E\text{-}STAR}_{95}\), \(\mathrm{E\text{-}Rachev}_{95,95}\), \(\mathrm{E\text{-}Rachev}_{50,95}\) \\
& CVaR benchmarks
& \(\mathrm{CVaR}_{95}\), \(\mathrm{STAR}_{95}\), \(\mathrm{Rachev}_{95,95}\), \(\mathrm{Rachev}_{50,95}\) \\
\bottomrule
\end{tabular}
\end{table}

	Each of the three tempered stable representations is evaluated with four portfolio objectives: minimum risk, STAR, symmetric Rachev, and asymmetric Rachev. Each EVaR portfolio is paired with the corresponding CVaR portfolio under the same representation. Normal EVaR portfolios are included as a thin-tailed benchmark. EW, Global Minimum Variance (MinVar), and Maximum Sharpe ratio (MaxSharpe) are included as standard allocation benchmarks.

	Portfolio construction uses rolling estimation windows. At each monthly rebalance date, the previous twelve months of daily returns are used to estimate the distributional parameters. The ICA models also require re-estimation of the mixing matrix. Algorithm~\ref{alg_evar_eval} is used to calculate portfolio EVaR, and the EVaR, E-STAR, and E-Rachev portfolio problems follow Section~\ref{sec_opt_objectives}. The feasible set imposes the long-only constraints \(w_i\ge0\) and the budget constraint \(\wt\boldsymbol{e}=1\), with no separate target-return constraint. The resulting positions remain fixed for one month and are updated at the next monthly rebalance date.

\subsubsection{Performance evaluation metrics}
\label{sec_evaluation_metrics}
	Portfolio performance is evaluated using standard return and risk measures. The reported measures include CAGR, annualized volatility, annualized Sharpe ratio, cumulative return, maximum drawdown, the CAGR-based Calmar ratio, and historical \(95\%\) VaR and CVaR. Skewness and excess kurtosis of realized portfolio returns are also reported. One-way portfolio turnover is computed at each rebalance date as \(\mathrm{TO}_t=\frac{1}{2}\sum_{i=1}^{N}|w_{i,t}-\widetilde{w}_{i,t^-}|\), where \(\widetilde{w}_{i,t^-}=w_{i,t-1}(1+R_{i,t^-})/\sum_{j=1}^{N}w_{j,t-1}(1+R_{j,t^-})\) denotes the drift-adjusted portfolio weight immediately before rebalancing and \(R_{i,t^-}\) is the simple return of asset \(i\) over the preceding holding period.

	A two-sided studentized circular block bootstrap following \cite{ledoit2008robust} tests Sharpe-ratio differences, with \(B=4999\) resamples and block length \(\lfloor T^{1/3}\rfloor\). The tables report bootstrap \(p\)-values with significance markers at the 10\%, 5\%, and 1\% levels. These tests assess uncertainty in the matched Sharpe-ratio differences. The economic comparisons also consider returns, drawdowns, turnover, regime performance, and transaction costs.

	The subperiod analysis divides the out-of-sample period into the eight calendar regimes shown in Table~\ref{tab:regime_definitions_us_spdr_sector_etf_yf_lb12_h1_non_rp_all}: the dot-com boom, the dot-com bust, the post-dot-com recovery, the pre-Global Financial Crisis (GFC) expansion, the GFC, the subsequent recovery, the COVID-and-inflation period, and the post-inflation expansion. Regime calculations use the overlap between these intervals and the available out-of-sample observations, so the first and final intervals are limited to the actual backtest period.

\begingroup
\begin{table}[!t]
\centering
\caption{Market-regime definitions used in the per-regime analysis. Dates are inclusive; regimes outside this universe's sample are omitted. (SECTOR, lookback=12m, holding=1m).}
\label{tab:regime_definitions_us_spdr_sector_etf_yf_lb12_h1_non_rp_all}
\small
\begin{tabular}{llll}
\toprule
Regime & Start & End & Description \\
\midrule
Dot-com boom & 1999-01-01 & 2000-03-31 & Late-stage technology run-up to the March 2000 peak \\
Dot-com bust & 2000-04-01 & 2003-03-31 & Technology bear market and 2001 recession \\
Post-dot-com recovery & 2003-04-01 & 2005-12-31 & Reflation and recovery ahead of the credit boom \\
Pre-GFC & 2006-01-01 & 2007-09-30 & Late-cycle credit boom \\
GFC & 2007-10-01 & 2009-06-30 & Global financial crisis \\
Recovery & 2009-07-01 & 2019-12-31 & Post-crisis bull market and low-volatility expansion \\
COVID + Inflation & 2020-01-01 & 2023-12-31 & COVID-19 crash, reopening, and the inflation/rate shock \\
Post-inflation expansion & 2024-01-01 & 2026-12-31 & Disinflation, monetary easing, and the AI-led equity expansion \\
\bottomrule
\end{tabular}
\end{table}
\endgroup

\subsection{Full-sample performance and matched risk-measure comparisons}
\label{sec:results_design}

	Table~\ref{tab:pair_design_us_spdr_sector_etf_yf_lb12_h1_non_rp_all} summarizes the main portfolio comparisons. In the EVaR--CVaR comparisons, the distributional model and multivariate representation are held fixed. Comparisons with the Normal benchmark may change both the tail model and the multivariate representation. Comparisons between ICA\(+\)NTS and MNTS change the dependence structure and the restrictions on the marginal parameters. These comparisons do not isolate a single modeling effect. Positive Sharpe differences favor the first specification. Positive differences in drawdown or turnover indicate deterioration.

\begingroup
\begin{table}[!t]
\centering
\caption{Experimental design. E denotes the EVaR-based variant and C the matched CVaR-based variant. The Normal specification is evaluated only for the entropic objectives. (SECTOR, lookback=12m, holding=1m).}
\label{tab:pair_design_us_spdr_sector_etf_yf_lb12_h1_non_rp_all}
\begin{tabular}{lcccc}
\toprule
 & Normal & MNTS projection & ICA$+$NTS & ICA$+$CTS \\
\midrule
Minimum risk & E & E $+$ C & E $+$ C & E $+$ C \\
STAR-type & E & E $+$ C & E $+$ C & E $+$ C \\
$\mathrm{Rachev}_{95,95}$ & E & E $+$ C & E $+$ C & E $+$ C \\
$\mathrm{Rachev}_{50,95}$ & E & E $+$ C & E $+$ C & E $+$ C \\
\bottomrule
\end{tabular}
\end{table}
\endgroup

	Table~\ref{tab:perf_full_us_spdr_sector_etf_yf_lb12_h1_non_rp_all} reports the performance of all thirty-one portfolios. The Normal minimum-EVaR portfolio and the minimum-variance benchmark are nearly identical, with gross Sharpe ratios of \(0.558\) and \(0.559\) and maximum drawdowns of \(37.27\%\) and \(37.36\%\), respectively. The ICA\(+\)NTS minimum-EVaR portfolio has the highest gross Sharpe ratio in the strategy set, \(0.616\), with a cumulative return of \(766.96\%\) and a maximum drawdown of \(38.30\%\). The symmetric MNTS E-Rachev portfolio has the highest cumulative return, \(976.15\%\), but its Sharpe ratio is \(0.474\) and its maximum drawdown is \(77.59\%\). As shown in Remark~\ref{cor_normal_ratio_equivalence}, Normal E-STAR and both Normal E-Rachev specifications coincide with MaxSharpe. Each has a cumulative return of \(604.55\%\), a Sharpe ratio of \(0.476\), and a maximum drawdown of \(51.74\%\).

\begingroup
\begin{table}[ht]
  \centering
  \footnotesize
  \caption{Portfolio performance, full backtest period (SECTOR, lookback=12m, holding=1m).}
  \label{tab:perf_full_us_spdr_sector_etf_yf_lb12_h1_non_rp_all}
  \resizebox{\linewidth}{!}{%
  \begin{tabular}{lccccccccccc}
    \toprule
    Portfolio & Cumul.\ return & CAGR & Ann.\ vol. & Sharpe & Sortino & Max DD & Calmar & VaR$_{95}$ & CVaR$_{95}$ & Skew. & Ex.\ kurt. \\
    \midrule
    EW & 735.31\% & 8.44\% & 18.16\% & 0.537 & 0.754 & 53.51\% & 0.158 & 1.71\% & 2.76\% & -0.248 & 11.555 \\
    MinVar & 509.64\% & 7.15\% & 14.15\% & 0.559 & 0.786 & 37.36\% & 0.191 & 1.31\% & 2.09\% & -0.224 & 14.275 \\
    MaxSharpe & 604.55\% & 7.74\% & 19.81\% & 0.476 & 0.668 & 51.74\% & 0.150 & 1.96\% & 2.99\% & -0.140 & 9.399 \\
    $\mathrm{EVaR}_{95}^{\mathrm{Normal}}$ & 506.49\% & 7.12\% & 14.13\% & 0.558 & 0.785 & 37.27\% & 0.191 & 1.31\% & 2.09\% & -0.232 & 13.993 \\
    $\mathrm{E\text{-}STAR}_{95}^{\mathrm{Normal}}$ & 604.55\% & 7.74\% & 19.81\% & 0.476 & 0.668 & 51.74\% & 0.150 & 1.96\% & 2.99\% & -0.140 & 9.399 \\
    $\mathrm{E\text{-}Rachev}_{95,95}^{\mathrm{Normal}}$ & 604.55\% & 7.74\% & 19.81\% & 0.476 & 0.668 & 51.74\% & 0.150 & 1.96\% & 2.99\% & -0.140 & 9.399 \\
    $\mathrm{E\text{-}Rachev}_{50,95}^{\mathrm{Normal}}$ & 604.55\% & 7.74\% & 19.81\% & 0.476 & 0.668 & 51.74\% & 0.150 & 1.96\% & 2.99\% & -0.140 & 9.399 \\
    $\mathrm{EVaR}_{95}^{\mathrm{MNTS}}$ & 499.57\% & 7.08\% & 14.13\% & 0.555 & 0.782 & 37.77\% & 0.187 & \textbf{1.30\%} & \textbf{2.08\%} & -0.182 & 12.398 \\
    $\mathrm{E\text{-}STAR}_{95}^{\mathrm{MNTS}}$ & 743.08\% & 8.48\% & 19.49\% & 0.515 & 0.723 & 49.04\% & 0.173 & 1.92\% & 2.94\% & -0.234 & 8.023 \\
    $\mathrm{E\text{-}Rachev}_{95,95}^{\mathrm{MNTS}}$ & \textbf{976.15\%} & \textbf{9.50\%} & 26.68\% & 0.474 & 0.681 & 77.59\% & 0.122 & 2.21\% & 3.88\% & 0.054 & 21.429 \\
    $\mathrm{E\text{-}Rachev}_{50,95}^{\mathrm{MNTS}}$ & 598.82\% & 7.71\% & 25.92\% & 0.417 & 0.592 & 77.49\% & 0.099 & 2.11\% & 3.81\% & -0.100 & 22.972 \\
    $\mathrm{CVaR}_{95}^{\mathrm{MNTS}}$ & 491.86\% & 7.02\% & \textbf{14.11\%} & 0.552 & 0.776 & \textbf{37.20\%} & 0.189 & 1.31\% & 2.08\% & -0.255 & 13.908 \\
    $\mathrm{STAR}_{95}^{\mathrm{MNTS}}$ & 726.70\% & 8.40\% & 19.34\% & 0.514 & 0.721 & 50.31\% & 0.167 & 1.91\% & 2.92\% & -0.218 & 8.059 \\
    $\mathrm{Rachev}_{95,95}^{\mathrm{MNTS}}$ & 925.24\% & 9.29\% & 25.99\% & 0.472 & 0.686 & 77.43\% & 0.120 & 2.23\% & 3.79\% & \textbf{0.320} & 17.311 \\
    $\mathrm{Rachev}_{50,95}^{\mathrm{MNTS}}$ & 442.98\% & 6.67\% & 20.93\% & 0.414 & 0.577 & 68.07\% & 0.098 & 1.91\% & 3.12\% & -0.278 & 18.340 \\
    $\mathrm{EVaR}_{95}^{\mathrm{NTS}}$ & 766.96\% & 8.60\% & 15.28\% & \textbf{0.616} & \textbf{0.880} & 38.30\% & \textbf{0.224} & 1.38\% & 2.21\% & 0.069 & 18.435 \\
    $\mathrm{E\text{-}STAR}_{95}^{\mathrm{NTS}}$ & 749.27\% & 8.51\% & 19.78\% & 0.512 & 0.719 & 52.75\% & 0.161 & 1.94\% & 3.00\% & -0.145 & 8.111 \\
    $\mathrm{E\text{-}Rachev}_{95,95}^{\mathrm{NTS}}$ & 818.73\% & 8.84\% & 21.21\% & 0.506 & 0.712 & 55.97\% & 0.158 & 2.00\% & 3.22\% & -0.254 & 9.507 \\
    $\mathrm{E\text{-}Rachev}_{50,95}^{\mathrm{NTS}}$ & 858.72\% & 9.01\% & 21.47\% & 0.510 & 0.726 & 58.68\% & 0.154 & 2.05\% & 3.21\% & 0.005 & 8.859 \\
    $\mathrm{CVaR}_{95}^{\mathrm{NTS}}$ & 495.43\% & 7.05\% & 14.22\% & 0.550 & 0.776 & 37.79\% & 0.187 & 1.32\% & 2.09\% & -0.142 & 14.636 \\
    $\mathrm{STAR}_{95}^{\mathrm{NTS}}$ & 718.09\% & 8.36\% & 19.55\% & 0.508 & 0.714 & 49.21\% & 0.170 & 1.95\% & 2.95\% & -0.206 & \textbf{7.806} \\
    $\mathrm{Rachev}_{95,95}^{\mathrm{NTS}}$ & 475.45\% & 6.91\% & 23.37\% & 0.403 & 0.570 & 69.89\% & 0.099 & 2.06\% & 3.50\% & 0.125 & 18.541 \\
    $\mathrm{Rachev}_{50,95}^{\mathrm{NTS}}$ & 667.47\% & 8.09\% & 20.48\% & 0.483 & 0.676 & 51.90\% & 0.156 & 1.99\% & 3.09\% & -0.304 & 8.351 \\
    $\mathrm{EVaR}_{95}^{\mathrm{CTS}}$ & 645.76\% & 7.97\% & 15.53\% & 0.572 & 0.813 & 39.21\% & 0.203 & 1.42\% & 2.27\% & 0.048 & 17.362 \\
    $\mathrm{E\text{-}STAR}_{95}^{\mathrm{CTS}}$ & 723.36\% & 8.38\% & 19.83\% & 0.505 & 0.709 & 53.38\% & 0.157 & 1.94\% & 3.01\% & -0.144 & 8.080 \\
    $\mathrm{E\text{-}Rachev}_{95,95}^{\mathrm{CTS}}$ & 519.84\% & 7.21\% & 20.93\% & 0.438 & 0.616 & 55.10\% & 0.131 & 2.00\% & 3.16\% & -0.213 & 9.784 \\
    $\mathrm{E\text{-}Rachev}_{50,95}^{\mathrm{CTS}}$ & 772.59\% & 8.62\% & 21.04\% & 0.498 & 0.710 & 60.58\% & 0.142 & 2.00\% & 3.14\% & 0.010 & 9.386 \\
    $\mathrm{CVaR}_{95}^{\mathrm{CTS}}$ & 517.18\% & 7.20\% & 14.23\% & 0.560 & 0.790 & 37.67\% & 0.191 & 1.31\% & 2.09\% & -0.144 & 14.636 \\
    $\mathrm{STAR}_{95}^{\mathrm{CTS}}$ & 707.96\% & 8.30\% & 19.55\% & 0.506 & 0.710 & 49.21\% & 0.169 & 1.95\% & 2.95\% & -0.206 & 7.809 \\
    $\mathrm{Rachev}_{95,95}^{\mathrm{CTS}}$ & 512.03\% & 7.16\% & 23.33\% & 0.413 & 0.584 & 70.11\% & 0.102 & 2.05\% & 3.50\% & 0.123 & 18.649 \\
    $\mathrm{Rachev}_{50,95}^{\mathrm{CTS}}$ & 717.30\% & 8.35\% & 20.43\% & 0.495 & 0.693 & 52.67\% & 0.159 & 1.97\% & 3.09\% & -0.305 & 8.425 \\
    \bottomrule
  \end{tabular}
  }
\end{table}
\endgroup

	Table~\ref{tab:benchmark_diff_sig_us_spdr_sector_etf_yf_lb12_h1_non_rp_all} shows that the ICA\(+\)NTS minimum-EVaR portfolio has a higher realized Sharpe ratio than EW, MinVar, and MaxSharpe in this sample. The other benchmark comparisons vary with the portfolio objective.

\begingroup
\begin{table}[!t]
\centering
\caption{Entropic portfolios versus the classical benchmarks. Each block reports the difference (entropic minus benchmark) in CAGR and annualised volatility (percentage points), Sharpe ratio, and Calmar ratio against the equal-weight, minimum-variance, and maximum-Sharpe portfolios. Significance stars on Sharpe differences follow the studentized circular block bootstrap test of Ledoit and Wolf (2008, Section 3), with $B = 4999$ resamples and block length $L = \lfloor T^{1/3} \rfloor$, at the 10\% ($^{*}$), 5\% ($^{**}$), and 1\% ($^{***}$) levels. LW $p$ is the two-sided $p$-value of the test. (SECTOR, lookback=12m, holding=1m).}
\label{tab:benchmark_diff_sig_us_spdr_sector_etf_yf_lb12_h1_non_rp_all}
\scriptsize
\setlength{\tabcolsep}{3pt}
\resizebox{\linewidth}{!}{%
\begin{tabular}{llrrrrrrrrrrrrrrr}
\toprule
 & & \multicolumn{5}{c}{vs EW} & \multicolumn{5}{c}{vs MinVar} & \multicolumn{5}{c}{vs MaxSharpe} \\
\cmidrule(lr){3-7}\cmidrule(lr){8-12}\cmidrule(lr){13-17}
Objective & Cell & $\Delta$CAGR & $\Delta$Vol & $\Delta$SR & LW $p$ & $\Delta$Calmar & $\Delta$CAGR & $\Delta$Vol & $\Delta$SR & LW $p$ & $\Delta$Calmar & $\Delta$CAGR & $\Delta$Vol & $\Delta$SR & LW $p$ & $\Delta$Calmar \\
\midrule
Minimum risk & MNTS & -1.36 & -4.02 & +0.017 & 0.849 & +0.030 & -0.07 & -0.02 & -0.004 & 0.873 & -0.004 & -0.66 & -5.68 & +0.079 & 0.513 & +0.038 \\
 & ICA$+$NTS & +0.15 & -2.87 & +0.079 & 0.438 & +0.067 & +1.45 & +1.13 & +0.058 & 0.320 & +0.033 & +0.86 & -4.53 & +0.141 & 0.268 & +0.075 \\
 & ICA$+$CTS & -0.47 & -2.63 & +0.034 & 0.737 & +0.046 & +0.83 & +1.37 & +0.013 & 0.840 & +0.012 & +0.23 & -4.28 & +0.096 & 0.465 & +0.054 \\
\addlinespace
STAR & MNTS & +0.04 & +1.33 & -0.022 & 0.835 & +0.015 & +1.33 & +5.34 & -0.043 & 0.716 & -0.018 & +0.74 & -0.32 & +0.040 & 0.164 & +0.023 \\
 & ICA$+$NTS & +0.07 & +1.63 & -0.025 & 0.814 & +0.004 & +1.36 & +5.63 & -0.047 & 0.713 & -0.030 & +0.77 & -0.03 & +0.036 & 0.380 & +0.012 \\
 & ICA$+$CTS & -0.06 & +1.67 & -0.032 & 0.772 & -0.001 & +1.24 & +5.67 & -0.053 & 0.673 & -0.034 & +0.64 & +0.02 & +0.030 & 0.533 & +0.007 \\
\addlinespace
$\mathrm{Rachev}_{95,95}$ & MNTS & +1.05 & +8.52 & -0.064 & 0.590 & -0.035 & +2.35 & +12.53 & -0.085 & 0.561 & -0.069 & +1.76 & +6.87 & -0.002 & 0.990 & -0.027 \\
 & ICA$+$NTS & +0.39 & +3.06 & -0.032 & 0.793 & +0.000 & +1.69 & +7.06 & -0.053 & 0.709 & -0.033 & +1.10 & +1.40 & +0.030 & 0.808 & +0.008 \\
 & ICA$+$CTS & -1.23 & +2.78 & -0.100 & 0.391 & -0.027 & +0.07 & +6.78 & -0.121 & 0.388 & -0.060 & -0.53 & +1.12 & -0.038 & 0.763 & -0.019 \\
\addlinespace
$\mathrm{Rachev}_{50,95}$ & MNTS & -0.74 & +7.76 & -0.121 & 0.280 & -0.058 & +0.56 & +11.76 & -0.142 & 0.304 & -0.092 & -0.03 & +6.10 & -0.059 & 0.664 & -0.050 \\
 & ICA$+$NTS & +0.57 & +3.31 & -0.028 & 0.825 & -0.004 & +1.87 & +7.31 & -0.049 & 0.745 & -0.038 & +1.27 & +1.65 & +0.034 & 0.760 & +0.004 \\
 & ICA$+$CTS & +0.18 & +2.89 & -0.039 & 0.754 & -0.015 & +1.48 & +6.89 & -0.060 & 0.687 & -0.049 & +0.88 & +1.23 & +0.023 & 0.843 & -0.007 \\
\bottomrule
\end{tabular}%
}
\end{table}
\endgroup

	Table~\ref{tab:matched_pairs_sig_us_spdr_sector_etf_yf_lb12_h1_non_rp_all} compares EVaR and CVaR under the same model and representation. Under MNTS, the four gross Sharpe differences are \(+0.003\), \(+0.001\), \(+0.002\), and \(+0.003\). The realized Sharpe ratios are very similar for EVaR and CVaR. Under ICA, seven of the eight differences are positive, and the CTS STAR difference is \(-0.001\). The largest NTS differences occur for minimum risk (\(+0.066\)) and symmetric Rachev (\(+0.103\)). For minimum risk, cumulative return increases from \(495.43\%\) under CVaR to \(766.96\%\) under EVaR, and maximum drawdown changes from \(37.79\%\) to \(38.30\%\). The CTS differences are smaller at \(+0.012\), \(-0.001\), \(+0.024\), and \(+0.003\).

\begingroup
\begin{table}[!t]
\centering
\caption{Matched comparison of entropic and classical portfolios with significance of the Sharpe difference. Significance stars on Sharpe differences follow the studentized circular block bootstrap test of Ledoit and Wolf (2008, Section 3), with $B = 4999$ resamples and block length $L = \lfloor T^{1/3} \rfloor$, at the 10\% ($^{*}$), 5\% ($^{**}$), and 1\% ($^{***}$) levels. LW $p$ is the two-sided $p$-value of the test. (SECTOR, lookback=12m, holding=1m).}
\label{tab:matched_pairs_sig_us_spdr_sector_etf_yf_lb12_h1_non_rp_all}
\scriptsize
\setlength{\tabcolsep}{3.4pt}
\resizebox{\linewidth}{!}{%
\begin{tabular}{llrrrrrrrrrr}
\toprule
 & & \multicolumn{4}{c}{EVaR-based} & \multicolumn{4}{c}{CVaR-based} & & \\
\cmidrule(lr){3-6}\cmidrule(lr){7-10}
Objective & Cell & CumRet & SR & MDD & TO & CumRet & SR & MDD & TO & $\Delta$SR & LW $p$ \\
\midrule
Minimum risk & MNTS & 499.57 & 0.555 & 37.77 & 9.52 & 491.86 & 0.552 & 37.20 & 7.80 & +0.003 & 0.868 \\
 & ICA$+$NTS & 766.96 & 0.616 & 38.30 & 21.67 & 495.43 & 0.550 & 37.79 & 9.76 & +0.066 & 0.216 \\
 & ICA$+$CTS & 645.76 & 0.572 & 39.21 & 27.66 & 517.18 & 0.560 & 37.67 & 10.97 & +0.012 & 0.847 \\
\addlinespace
STAR & MNTS & 743.08 & 0.515 & 49.04 & 32.03 & 726.70 & 0.514 & 50.31 & 31.71 & +0.001 & 0.903 \\
 & ICA$+$NTS & 749.27 & 0.512 & 52.75 & 37.59 & 718.09 & 0.508 & 49.21 & 32.21 & +0.004 & 0.915 \\
 & ICA$+$CTS & 723.36 & 0.505 & 53.38 & 40.73 & 707.96 & 0.506 & 49.21 & 32.34 & -0.001 & 0.989 \\
\addlinespace
$\mathrm{Rachev}_{95,95}$ & MNTS & 976.15 & 0.474 & 77.59 & 33.01 & 925.24 & 0.472 & 77.43 & 29.65 & +0.002 & 0.976 \\
 & ICA$+$NTS & 818.73 & 0.506 & 55.97 & 62.26 & 475.45 & 0.403 & 69.89 & 53.23 & +0.103 & 0.314 \\
 & ICA$+$CTS & 519.84 & 0.438 & 55.10 & 64.74 & 512.03 & 0.413 & 70.11 & 52.07 & +0.024 & 0.807 \\
\addlinespace
$\mathrm{Rachev}_{50,95}$ & MNTS & 598.82 & 0.417 & 77.49 & 32.05 & 442.98 & 0.414 & 68.07 & 36.42 & +0.003 & 0.977 \\
 & ICA$+$NTS & 858.72 & 0.510 & 58.68 & 59.18 & 667.47 & 0.483 & 51.90 & 42.42 & +0.027 & 0.803 \\
 & ICA$+$CTS & 772.59 & 0.498 & 60.58 & 63.59 & 717.30 & 0.495 & 52.67 & 42.51 & +0.003 & 0.976 \\
\bottomrule
\end{tabular}%
}
\end{table}
\endgroup

	Table~\ref{tab:normal_benchmark_sig_us_spdr_sector_etf_yf_lb12_h1_non_rp_all} compares the tempered stable entropic portfolios with the matched Normal benchmark. ICA\(+\)NTS has positive gross Sharpe differences for all four objectives (\(+0.058\), \(+0.036\), \(+0.030\), and \(+0.034\)). After 25 basis points of transaction costs, the differences remain positive at minimum risk (\(+0.030\)) and STAR (\(+0.027\)). MNTS is close to Normal at minimum risk (\(-0.003\) gross), higher at STAR (\(+0.040\)), and lower in the two Rachev portfolios (\(-0.002\) and \(-0.059\)). The ICA\(+\)CTS results are also mixed. In this sample, the larger EVaR--CVaR differences occur mainly under ICA. The larger differences relative to the Normal benchmark occur mainly for minimum risk and STAR.

\begingroup
\begin{table}[!t]
\centering
\caption{Entropic portfolios relative to the matched Gaussian EVaR benchmark. The table reports the Gaussian gross and net Sharpe ratios and the Sharpe differences of the tempered-stable specifications, gross and net of a 25 basis point transaction-cost assumption. Significance stars on Sharpe differences follow the studentized circular block bootstrap test of Ledoit and Wolf (2008, Section 3), with $B = 4999$ resamples and block length $L = \lfloor T^{1/3} \rfloor$, at the 10\% ($^{*}$), 5\% ($^{**}$), and 1\% ($^{***}$) levels. LW $p$ is the two-sided $p$-value of the test. (SECTOR, lookback=12m, holding=1m).}
\label{tab:normal_benchmark_sig_us_spdr_sector_etf_yf_lb12_h1_non_rp_all}
\scriptsize
\setlength{\tabcolsep}{4.2pt}
\resizebox{\linewidth}{!}{%
\begin{tabular}{lrr|rrrr|rrrr|rrrr}
\toprule
 & \multicolumn{2}{c|}{Normal benchmark} & \multicolumn{4}{c|}{MNTS} & \multicolumn{4}{c|}{ICA$+$NTS} & \multicolumn{4}{c}{ICA$+$CTS} \\
\cmidrule(lr){2-3}\cmidrule(lr){4-7}\cmidrule(lr){8-11}\cmidrule(lr){12-15}
Objective & Gross SR & Net SR & $\Delta$ gross & LW $p$ & $\Delta$ net & LW $p$ & $\Delta$ gross & LW $p$ & $\Delta$ net & LW $p$ & $\Delta$ gross & LW $p$ & $\Delta$ net & LW $p$ \\
\midrule
Minimum risk & 0.558 & 0.543 & -0.003 & 0.893 & -0.009 & 0.719 & +0.058 & 0.316 & +0.030 & 0.590 & +0.014 & 0.826 & -0.025 & 0.702 \\
STAR & 0.476 & 0.427 & +0.040 & 0.164 & +0.039 & 0.184 & +0.036 & 0.380 & +0.027 & 0.508 & +0.030 & 0.533 & +0.016 & 0.737 \\
$\mathrm{Rachev}_{95,95}$ & 0.476 & 0.427 & -0.002 & 0.990 & +0.009 & 0.949 & +0.030 & 0.808 & -0.010 & 0.934 & -0.038 & 0.763 & -0.083 & 0.505 \\
$\mathrm{Rachev}_{50,95}$ & 0.476 & 0.427 & -0.059 & 0.664 & -0.048 & 0.724 & +0.034 & 0.760 & -0.001 & 0.993 & +0.023 & 0.843 & -0.020 & 0.858 \\
\bottomrule
\end{tabular}%
}
\end{table}
\endgroup

\subsection{Distributional representation, tail family, and implementation}
\label{sec:results_route}

	Table~\ref{tab:route_tail_sig_us_spdr_sector_etf_yf_lb12_h1_non_rp_all} compares CTS with NTS within ICA as well as ICA\(+\)NTS with MNTS. The ICA\(+\)NTS--MNTS comparison does not isolate only the dependence model. MNTS uses a shared subordinator with common \(\alpha,\theta\). ICA\(+\)NTS allows component-specific \(\alpha_i,\theta_i\). The comparison includes both the dependence structure and the restrictions on the marginal parameters.

	Under EVaR, the ICA\(+\)NTS--MNTS Sharpe differences are positive at minimum risk (\(+0.062\)) and for the symmetric and asymmetric Rachev portfolios (\(+0.032\) and \(+0.093\)). The STAR difference is close to zero and slightly negative. Under CVaR, the corresponding differences are \(-0.002\) at minimum risk, \(-0.069\) for symmetric Rachev, and \(+0.069\) for asymmetric Rachev. For the two EVaR Rachev portfolios, the ICA\(+\)NTS portfolios also reduce maximum drawdown by \(21.61\) and \(18.81\) percentage points relative to MNTS.

	Within ICA, the absolute CTS--NTS Sharpe difference is larger under EVaR in three of the four objectives: \(-0.044\) versus \(+0.009\) at minimum risk, \(-0.068\) versus \(+0.010\) for symmetric Rachev, and \(-0.007\) versus \(-0.002\) for STAR. The asymmetric Rachev differences have similar magnitudes. NTS has the higher Sharpe ratio in the main EVaR comparisons in this sample, but this result does not imply a general ranking of NTS and CTS.

\begingroup
\begin{table}[!t]
\centering
\caption{Tail-family and distributional-route effects. The CTS--NTS columns compare ICA marginal families. The remaining columns report ICA$+$NTS minus MNTS, holding the objective and risk measure fixed. Return, drawdown, and turnover differences are in percentage points. Net Sharpe differences use a 25 basis point proportional transaction-cost assumption. Significance stars on Sharpe differences follow the studentized circular block bootstrap test of Ledoit and Wolf (2008, Section 3), with $B = 4999$ resamples and block length $L = \lfloor T^{1/3} \rfloor$, at the 10\% ($^{*}$), 5\% ($^{**}$), and 1\% ($^{***}$) levels. LW $p$ is the two-sided $p$-value of the test. (SECTOR, lookback=12m, holding=1m).}
\label{tab:route_tail_sig_us_spdr_sector_etf_yf_lb12_h1_non_rp_all}
\scriptsize
\setlength{\tabcolsep}{3.2pt}
\resizebox{\linewidth}{!}{%
\begin{tabular}{llrrrr|rrrrrrr}
\toprule
 & & \multicolumn{4}{c|}{CTS $-$ NTS within ICA} & \multicolumn{7}{c}{ICA$+$NTS $-$ MNTS} \\
\cmidrule(lr){3-6}\cmidrule(lr){7-13}
Objective & Measure & Gross $\Delta$SR & LW $p$ & Net $\Delta$SR & LW $p$ & $\Delta$CumRet & $\Delta$SR & LW $p$ & $\Delta$MDD & $\Delta$TO & Net $\Delta$SR & LW $p$ \\
\midrule
Minimum risk & EVaR & -0.044 & 0.254 & -0.055 & 0.158 & +267.38 & +0.062 & 0.299 & +0.52 & +12.14 & +0.039 & 0.507 \\
 & CVaR & +0.009 & 0.113 & +0.007 & 0.240 & +3.57 & -0.002 & 0.920 & +0.59 & +1.96 & -0.006 & 0.730 \\
\addlinespace
STAR & EVaR & -0.007 & 0.704 & -0.011 & 0.522 & +6.19 & -0.003 & 0.923 & +3.71 & +5.56 & -0.011 & 0.762 \\
 & CVaR & -0.002 & 0.444 & -0.003 & 0.438 & -8.60 & -0.006 & 0.674 & -1.10 & +0.49 & -0.006 & 0.666 \\
\addlinespace
$\mathrm{Rachev}_{95,95}$ & EVaR & -0.068 & 0.220 & -0.073 & 0.189 & -157.43 & +0.032 & 0.824 & -21.61 & +29.25 & -0.019 & 0.894 \\
 & CVaR & +0.010 & 0.543 & +0.012 & 0.491 & -449.79 & -0.069 & 0.576 & -7.54 & +23.58 & -0.103 & 0.411 \\
\addlinespace
$\mathrm{Rachev}_{50,95}$ & EVaR & -0.011 & 0.837 & -0.019 & 0.731 & +259.90 & +0.093 & 0.505 & -18.81 & +27.12 & +0.047 & 0.731 \\
 & CVaR & +0.012 & 0.505 & +0.012 & 0.520 & +224.49 & +0.069 & 0.546 & -16.17 & +5.99 & +0.059 & 0.609 \\
\bottomrule
\end{tabular}%
}
\end{table}
\endgroup

	Table~\ref{tab:turnover_summary_us_spdr_sector_etf_yf_lb12_h1_non_rp_all} reports portfolio turnover. EW, MinVar, Normal minimum-EVaR, and MNTS minimum-EVaR have relatively low mean turnover between \(1.35\%\) and \(9.52\%\). Mean turnover increases to \(21.67\%\) for ICA\(+\)NTS minimum-EVaR and \(27.66\%\) for ICA\(+\)CTS minimum-EVaR. The ICA-based Rachev portfolios have the highest mean turnover, from \(59.18\%\) to \(64.74\%\), and also have high median turnover. The MNTS Rachev portfolios trade differently. The symmetric and asymmetric E-Rachev portfolios have mean turnovers of \(33.01\%\) and \(32.05\%\), but their medians are \(13.90\%\) and \(17.97\%\). The classical symmetric MNTS Rachev portfolio has a median turnover of zero. Turnover alone does not show that the optimizer is unstable. It does show that the ICA and MNTS portfolios can require substantially different amounts of rebalancing.

\begingroup
\begin{table}[ht]
  \centering
  \caption{Per-rebalance turnover statistics across portfolios. Turnover is the one-way fraction of NAV traded against the drift-adjusted weight from the previous holding period.  Lower turnover implies lower transaction cost for any cost rate. (SECTOR, lookback=12m, holding=1m).}
  \label{tab:turnover_summary_us_spdr_sector_etf_yf_lb12_h1_non_rp_all}
  \begin{tabular}{lccc}
    \toprule
    Portfolio & mean & median & std \\
    \midrule
    EW & 1.35\% & 1.13\% & 0.91\% \\
    MinVar & 6.54\% & 5.33\% & 5.14\% \\
    MaxSharpe & 31.69\% & 26.67\% & 25.36\% \\
    $\mathrm{EVaR}_{95}^{\mathrm{Normal}}$ & 6.93\% & 5.86\% & 5.42\% \\
    $\mathrm{E\text{-}STAR}_{95}^{\mathrm{Normal}}$ & 31.69\% & 26.67\% & 25.36\% \\
    $\mathrm{E\text{-}Rachev}_{95,95}^{\mathrm{Normal}}$ & 31.69\% & 26.67\% & 25.36\% \\
    $\mathrm{E\text{-}Rachev}_{50,95}^{\mathrm{Normal}}$ & 31.69\% & 26.67\% & 25.36\% \\
    $\mathrm{EVaR}_{95}^{\mathrm{MNTS}}$ & 9.52\% & 8.04\% & 6.64\% \\
    $\mathrm{E\text{-}STAR}_{95}^{\mathrm{MNTS}}$ & 32.03\% & 28.98\% & 24.85\% \\
    $\mathrm{E\text{-}Rachev}_{95,95}^{\mathrm{MNTS}}$ & 33.01\% & 13.90\% & 39.12\% \\
    $\mathrm{E\text{-}Rachev}_{50,95}^{\mathrm{MNTS}}$ & 32.05\% & 17.97\% & 36.99\% \\
    $\mathrm{CVaR}_{95}^{\mathrm{MNTS}}$ & 7.80\% & 6.63\% & 6.17\% \\
    $\mathrm{STAR}_{95}^{\mathrm{MNTS}}$ & 31.71\% & 27.34\% & 24.66\% \\
    $\mathrm{Rachev}_{95,95}^{\mathrm{MNTS}}$ & 29.65\% & 0.00\% & 39.38\% \\
    $\mathrm{Rachev}_{50,95}^{\mathrm{MNTS}}$ & 36.42\% & 31.49\% & 32.32\% \\
    $\mathrm{EVaR}_{95}^{\mathrm{NTS}}$ & 21.67\% & 16.71\% & 20.34\% \\
    $\mathrm{E\text{-}STAR}_{95}^{\mathrm{NTS}}$ & 37.59\% & 32.56\% & 30.32\% \\
    $\mathrm{E\text{-}Rachev}_{95,95}^{\mathrm{NTS}}$ & 62.26\% & 72.44\% & 37.83\% \\
    $\mathrm{E\text{-}Rachev}_{50,95}^{\mathrm{NTS}}$ & 59.18\% & 64.81\% & 38.45\% \\
    $\mathrm{CVaR}_{95}^{\mathrm{NTS}}$ & 9.76\% & 8.35\% & 7.34\% \\
    $\mathrm{STAR}_{95}^{\mathrm{NTS}}$ & 32.21\% & 26.48\% & 25.64\% \\
    $\mathrm{Rachev}_{95,95}^{\mathrm{NTS}}$ & 53.23\% & 54.82\% & 38.74\% \\
    $\mathrm{Rachev}_{50,95}^{\mathrm{NTS}}$ & 42.42\% & 37.95\% & 34.95\% \\
    $\mathrm{EVaR}_{95}^{\mathrm{CTS}}$ & 27.66\% & 20.91\% & 24.22\% \\
    $\mathrm{E\text{-}STAR}_{95}^{\mathrm{CTS}}$ & 40.73\% & 36.09\% & 32.84\% \\
    $\mathrm{E\text{-}Rachev}_{95,95}^{\mathrm{CTS}}$ & 64.74\% & 79.86\% & 38.12\% \\
    $\mathrm{E\text{-}Rachev}_{50,95}^{\mathrm{CTS}}$ & 63.59\% & 74.16\% & 36.69\% \\
    $\mathrm{CVaR}_{95}^{\mathrm{CTS}}$ & 10.97\% & 9.48\% & 8.67\% \\
    $\mathrm{STAR}_{95}^{\mathrm{CTS}}$ & 32.34\% & 27.07\% & 25.67\% \\
    $\mathrm{Rachev}_{95,95}^{\mathrm{CTS}}$ & 52.07\% & 55.05\% & 37.64\% \\
    $\mathrm{Rachev}_{50,95}^{\mathrm{CTS}}$ & 42.51\% & 37.15\% & 33.22\% \\
    \bottomrule
  \end{tabular}
\end{table}
\endgroup

\subsection{Transaction costs and market regimes}
\label{sec:results_robustness}

	Table~\ref{tab:cost_premium_sig_us_spdr_sector_etf_yf_lb12_h1_non_rp_all} reports matched EVaR--CVaR Sharpe differences at transaction costs of 0, 5, 10, and 25 basis points. Table~\ref{tab:cost_impact_combined_us_spdr_sector_etf_yf_lb12_h1_non_rp_all} in the appendix reports the complete gross and net results. For ICA\(+\)NTS, the minimum-risk Sharpe difference decreases from \(+0.066\) to \(+0.044\) at 25 basis points, and the symmetric Rachev difference decreases from \(+0.103\) to \(+0.083\). The asymmetric NTS difference also remains positive but decreases from \(+0.027\) to \(+0.006\). The CTS differences decrease to zero or below at 10 to 25 basis points, and the MNTS differences remain small. The ICA\(+\)NTS minimum-EVaR portfolio has the highest net Sharpe ratio at each reported positive cost rate, with values of \(0.608\), \(0.599\), and \(0.573\) at 5, 10, and 25 basis points.

\begingroup
\begin{table}[!t]
\centering
\caption{Transaction-cost path of the EVaR--CVaR Sharpe premium. Each entry is the Sharpe ratio of the EVaR-based portfolio minus that of the matched CVaR-based portfolio. The zero-cost column is the gross difference. Significance stars on Sharpe differences follow the studentized circular block bootstrap test of Ledoit and Wolf (2008, Section 3), with $B = 4999$ resamples and block length $L = \lfloor T^{1/3} \rfloor$, at the 10\% ($^{*}$), 5\% ($^{**}$), and 1\% ($^{***}$) levels. LW $p$ is the two-sided $p$-value of the test. (SECTOR, lookback=12m, holding=1m).}
\label{tab:cost_premium_sig_us_spdr_sector_etf_yf_lb12_h1_non_rp_all}
\scriptsize
\setlength{\tabcolsep}{4pt}
\begin{tabular}{llrrrrrrrr}
\toprule
Objective & Cell & 0 bps & LW $p$ & 5 bps & LW $p$ & 10 bps & LW $p$ & 25 bps & LW $p$ \\
\midrule
Minimum risk & MNTS & +0.003 & 0.868 & +0.002 & 0.900 & +0.001 & 0.935 & -0.001 & 0.959 \\
 & ICA$+$NTS & +0.066 & 0.216 & +0.061 & 0.249 & +0.057 & 0.288 & +0.044 & 0.409 \\
 & ICA$+$CTS & +0.012 & 0.847 & +0.006 & 0.924 & -0.000 & 0.998 & -0.018 & 0.763 \\
\addlinespace
STAR & MNTS & +0.001 & 0.903 & +0.001 & 0.905 & +0.001 & 0.907 & +0.001 & 0.913 \\
 & ICA$+$NTS & +0.004 & 0.915 & +0.002 & 0.951 & +0.001 & 0.989 & -0.004 & 0.907 \\
 & ICA$+$CTS & -0.001 & 0.989 & -0.003 & 0.944 & -0.005 & 0.897 & -0.013 & 0.766 \\
\addlinespace
$\mathrm{Rachev}_{95,95}$ & MNTS & +0.002 & 0.976 & +0.001 & 0.985 & +0.001 & 0.991 & -0.001 & 0.990 \\
 & ICA$+$NTS & +0.103 & 0.314 & +0.099 & 0.332 & +0.095 & 0.347 & +0.083 & 0.409 \\
 & ICA$+$CTS & +0.024 & 0.807 & +0.019 & 0.849 & +0.014 & 0.891 & -0.001 & 0.988 \\
\addlinespace
$\mathrm{Rachev}_{50,95}$ & MNTS & +0.003 & 0.977 & +0.006 & 0.953 & +0.009 & 0.927 & +0.018 & 0.854 \\
 & ICA$+$NTS & +0.027 & 0.803 & +0.023 & 0.832 & +0.019 & 0.862 & +0.006 & 0.956 \\
 & ICA$+$CTS & +0.003 & 0.976 & -0.002 & 0.984 & -0.008 & 0.942 & -0.025 & 0.821 \\
\bottomrule
\end{tabular}
\end{table}
\endgroup

	The regime results show that the full-sample differences change over time. Table~\ref{tab:regime_rachev_us_spdr_sector_etf_yf_lb12_h1_non_rp_all} reports selected Rachev results. Table~\ref{tab:perf_subperiods_combined_us_spdr_sector_etf_yf_lb12_h1_non_rp_all} in the appendix reports all regime results. For ICA\(+\)NTS minimum risk, the EVaR portfolio has a higher cumulative return than the matched CVaR portfolio in six of the eight regimes. During the GFC, the NTS and CTS minimum-EVaR portfolios lose \(12.18\%\) and \(11.70\%\). The matched NTS minimum-CVaR portfolio loses \(17.61\%\), MinVar loses \(18.60\%\), and EW loses \(33.77\%\). The NTS minimum-EVaR portfolio gains \(364.59\%\) in the subsequent recovery and \(58.73\%\) in the COVID-and-inflation period, compared with \(289.84\%\) and \(26.96\%\) for minimum-CVaR. In the post-inflation expansion, minimum-EVaR gains \(8.87\%\) and minimum-CVaR gains \(19.66\%\).

	The Rachev portfolios show larger differences across regimes. In the dot-com bust, the symmetric MNTS E-Rachev portfolio gains \(1.18\%\). The corresponding ICA\(+\)NTS and ICA\(+\)CTS E-Rachev portfolios lose \(39.30\%\) and \(33.11\%\). All selected Rachev portfolios lose during the GFC, with cumulative returns between \(-26.53\%\) and \(-52.36\%\). The symmetric ICA\(+\)NTS E-Rachev portfolio gains \(374.32\%\) in the recovery and \(154.82\%\) in the COVID-and-inflation period. Its matched classical Rachev portfolio gains \(188.03\%\) and \(148.51\%\), respectively. The ranking changes again in the post-inflation expansion.

\begingroup
\begin{table}[!t]
\centering
\caption{Selected Rachev portfolios across market regimes. Entries are cumulative returns in percent. (SECTOR, lookback=12m, holding=1m).}
\label{tab:regime_rachev_us_spdr_sector_etf_yf_lb12_h1_non_rp_all}
\scriptsize
\setlength{\tabcolsep}{3.5pt}
\resizebox{\linewidth}{!}{%
\begin{tabular}{lrrrrrrrr}
\toprule
Portfolio & Dot-com boom & Dot-com bust & Post-dot-com recovery & Pre-GFC & GFC & Recovery & COVID + Inflation & Post-inflation expansion \\
\midrule
Normal E-Rachev & 2.86 & -39.69 & 79.84 & 13.12 & -28.59 & 202.97 & 78.28 & 44.74 \\
MNTS E-Rachev & 22.06 & 1.18 & 55.84 & 7.59 & -52.36 & 280.28 & 115.44 & 33.15 \\
ICA+NTS E-Rachev & -1.16 & -39.30 & 47.17 & 4.88 & -37.97 & 374.32 & 154.82 & 32.34 \\
ICA+NTS Rachev & -4.45 & -37.04 & 51.89 & 24.21 & -50.94 & 188.03 & 148.51 & 44.38 \\
ICA+CTS E-Rachev & -4.99 & -33.11 & 49.69 & 8.13 & -39.03 & 256.80 & 112.60 & 30.29 \\
ICA+CTS Rachev & -4.29 & -36.98 & 47.19 & 24.23 & -51.52 & 199.52 & 149.95 & 52.90 \\
ICA+CTS asymmetric E-Rachev & 0.42 & -28.64 & 78.23 & 7.92 & -26.53 & 247.15 & 81.41 & 36.81 \\
\bottomrule
\end{tabular}%
}
\end{table}
\endgroup

\section{Conclusion}
\label{sec_conclusion}

	We develop parametric EVaR portfolio optimization for tempered stable L\'evy returns. Under the MNTS and ICA representations, the fitted parameters and portfolio weights determine the portfolio cumulant-generating function and its admissible moment-generating-function domain. This allows portfolio EVaR to be calculated across candidate weights using the same fitted parameters. We use this approach to construct minimum-EVaR, E-STAR, and E-Rachev portfolios.

	Portfolio outcomes differ across the risk measure, multivariate representation, tail family, and portfolio objective. In this sample, the ICA\(+\)NTS minimum-EVaR portfolio has the highest realized Sharpe ratio among the strategies considered. The EVaR--CVaR differences are generally small under MNTS and more pronounced under the ICA specifications. Turnover, transaction costs, and market regimes also affect the realized portfolio results. The findings are specific to the asset universe, sample period, and backtest design and do not imply that EVaR always outperforms CVaR or that one tempered stable model is always better than another.


\appendix
\section{Supplementary performance tables}
\label{app_tables}
	This appendix reports the complete strategy-level statistics that complement the contrast tables of Section~\ref{sec_result}: the transaction-cost impact of every strategy at each cost rate (Table~\ref{tab:cost_impact_combined_us_spdr_sector_etf_yf_lb12_h1_non_rp_all}) and the regime-level performance of every strategy (Table~\ref{tab:perf_subperiods_combined_us_spdr_sector_etf_yf_lb12_h1_non_rp_all}).

\begingroup
\footnotesize
\begin{longtable}{llrrrrrrrrr}
  \caption{Cost impact across rates (SECTOR, lookback=12m, holding=1m).  For each cost rate, the table reports gross, net, and $\Delta = $ net $-$ gross for the headline metrics.} \label{tab:cost_impact_combined_us_spdr_sector_etf_yf_lb12_h1_non_rp_all} \\
  \toprule
   &  & \multicolumn{3}{c}{Cumul.\ return} & \multicolumn{3}{c}{Sharpe} & \multicolumn{3}{c}{Calmar} \\
  \cmidrule(lr){3-5} \cmidrule(lr){6-8} \cmidrule(lr){9-11}
  Cost rate & Portfolio & gross & net & $\Delta$ & gross & net & $\Delta$ & gross & net & $\Delta$ \\
  \midrule
\endfirsthead
  \multicolumn{11}{l}{\textit{Table~\ref{tab:cost_impact_combined_us_spdr_sector_etf_yf_lb12_h1_non_rp_all} (continued)}} \\
  \toprule
   &  & \multicolumn{3}{c}{Cumul.\ return} & \multicolumn{3}{c}{Sharpe} & \multicolumn{3}{c}{Calmar} \\
  \cmidrule(lr){3-5} \cmidrule(lr){6-8} \cmidrule(lr){9-11}
  Cost rate & Portfolio & gross & net & $\Delta$ & gross & net & $\Delta$ & gross & net & $\Delta$ \\
  \midrule
\endhead
  \bottomrule
\endlastfoot
  \multirow{31}{*}{5 bps} & EW & 735.31\% & 733.55\% & -1.77\% & 0.537 & 0.537 & -0.000 & 0.158 & 0.158 & -0.000 \\
   & MinVar & 509.64\% & 503.39\% & -6.25\% & 0.559 & 0.556 & -0.003 & 0.191 & 0.190 & -0.001 \\
   & MaxSharpe & 604.55\% & 570.28\% & -34.27\% & 0.476 & 0.466 & -0.010 & 0.150 & 0.145 & -0.004 \\
   & $\mathrm{EVaR}_{95}^{\mathrm{Normal}}$ & 506.49\% & 499.91\% & -6.58\% & 0.558 & 0.555 & -0.003 & 0.191 & 0.190 & -0.001 \\
   & $\mathrm{E\text{-}STAR}_{95}^{\mathrm{Normal}}$ & 604.55\% & 570.28\% & -34.27\% & 0.476 & 0.466 & -0.010 & 0.150 & 0.145 & -0.004 \\
   & $\mathrm{E\text{-}Rachev}_{95,95}^{\mathrm{Normal}}$ & 604.55\% & 570.28\% & -34.27\% & 0.476 & 0.466 & -0.010 & 0.150 & 0.145 & -0.004 \\
   & $\mathrm{E\text{-}Rachev}_{50,95}^{\mathrm{Normal}}$ & 604.55\% & 570.28\% & -34.27\% & 0.476 & 0.466 & -0.010 & 0.150 & 0.145 & -0.004 \\
   & $\mathrm{EVaR}_{95}^{\mathrm{MNTS}}$ & 499.57\% & 490.66\% & -8.91\% & 0.555 & 0.551 & -0.004 & 0.187 & 0.185 & -0.002 \\
\end{longtable}
\normalsize
\endgroup

\begin{landscape}
\begingroup
\setlength{\tabcolsep}{3pt}
\renewcommand{\arraystretch}{0.92}
\scriptsize
\begin{longtable}{llccccccccccc}
  \caption{Portfolio performance across regimes (SECTOR, lookback=12m, holding=1m).} \label{tab:perf_subperiods_combined_us_spdr_sector_etf_yf_lb12_h1_non_rp_all} \\
  \toprule
  Regime & Portfolio & Cumul.\ return & CAGR & Ann.\ vol. & Sharpe & Sortino & Max DD & Calmar & VaR$_{95}$ & CVaR$_{95}$ & Skew. & Ex.\ kurt. \\
  \midrule
\endfirsthead
  \multicolumn{13}{l}{\textit{Table~\ref{tab:perf_subperiods_combined_us_spdr_sector_etf_yf_lb12_h1_non_rp_all} (continued)}} \\
  \toprule
  Regime & Portfolio & Cumul.\ return & CAGR & Ann.\ vol. & Sharpe & Sortino & Max DD & Calmar & VaR$_{95}$ & CVaR$_{95}$ & Skew. & Ex.\ kurt. \\
  \midrule
\endhead
  \bottomrule
\endlastfoot
  \multirow{31}{*}{Dot-com boom} & EW & -0.86\% & -3.41\% & 22.52\% & -0.044 & -0.066 & 13.64\% & -0.250 & 2.16\% & 2.65\% & 0.647 & 2.128 \\
   & MinVar & -2.00\% & -7.76\% & 21.43\% & -0.272 & -0.401 & 14.43\% & -0.538 & 2.07\% & 2.47\% & 0.454 & 0.876 \\
   & MaxSharpe & 2.86\% & 11.94\% & 24.90\% & 0.576 & 0.802 & 10.32\% & 1.157 & 2.82\% & 3.20\% & -0.296 & -0.301 \\
   & $\mathrm{EVaR}_{95}^{\mathrm{Normal}}$ & -1.68\% & -6.57\% & 21.10\% & -0.218 & -0.322 & 14.26\% & -0.460 & \textbf{1.96\%} & 2.47\% & 0.420 & 0.954 \\
   & $\mathrm{E\text{-}STAR}_{95}^{\mathrm{Normal}}$ & 2.86\% & 11.94\% & 24.90\% & 0.576 & 0.802 & 10.32\% & 1.157 & 2.82\% & 3.20\% & -0.296 & -0.301 \\
   & $\mathrm{E\text{-}Rachev}_{95,95}^{\mathrm{Normal}}$ & 2.86\% & 11.94\% & 24.90\% & 0.576 & 0.802 & 10.32\% & 1.157 & 2.82\% & 3.20\% & -0.296 & -0.301 \\
   & $\mathrm{E\text{-}Rachev}_{50,95}^{\mathrm{Normal}}$ & 2.86\% & 11.94\% & 24.90\% & 0.576 & 0.802 & 10.32\% & 1.157 & 2.82\% & 3.20\% & -0.296 & -0.301 \\
   & $\mathrm{EVaR}_{95}^{\mathrm{MNTS}}$ & -3.17\% & -12.09\% & 22.14\% & -0.473 & -0.695 & 17.11\% & -0.707 & 2.05\% & \textbf{2.47\%} & 0.513 & 1.096 \\
\end{longtable}
\normalsize
\endgroup
\end{landscape}

\bibliographystyle{plainnat}
\bibliography{TSEVaRPortfolio}

\end{document}